\documentclass[preprint,12pt,review]{elsarticle}
\usepackage[T1]{fontenc}

\DeclareUnicodeCharacter{2032}{\ensuremath{\prime}}
\usepackage{amsmath}
\usepackage{amssymb}
\usepackage{color}
\usepackage{titlesec}
\usepackage{booktabs, threeparttable, stackengine}
\usepackage{tabu}
\usepackage{tabularx}
\usepackage{graphicx}    
\usepackage{subfigure}   
\usepackage{float} 
\usepackage{placeins}  
\usepackage{multirow}
\usepackage{array}
\usepackage{gensymb}
\usepackage{colortbl}
\usepackage{enumitem}
\usepackage{xcolor}
\newcolumntype{L}[1]{>{\raggedright\arraybackslash}p{#1}}
\setstackEOL{\#}
\setstackgap{L}{12pt}
\usepackage[font=small]{caption}
\extrafloats{100}
\usepackage{makecell}
\usepackage{mathrsfs}  
\usepackage{hyperref}

\titleformat{\paragraph}
{\normalfont\normalsize\bfseries}{\theparagraph}{1em}{}
\titlespacing*{\paragraph}
{0pt}{3.25ex plus 1ex minus .2ex}{1.5ex plus .2ex}

\biboptions{sort&compress}

\journal{International Journal of Heat and Mass Transfer}

\begin{document}
\begin{frontmatter}
\title{Radiative equilibrium boundary condition and correlation analysis on catalytic surfaces in DSMC}

\author[inst1]{Youngil Ko}
\author[inst1]{Eunji Jun\corref{cor}}
\affiliation[inst1]{organization={Korea Advanced Institute of Science and Technology},
            postcode ={34141}, 
            state={Daejeon},
            country={Republic of Korea}}
\ead{eunji.jun@kaist.ac.kr}
\cortext[cor]{Corresponding author}

    \begin{abstract}

    This study integrates radiative equilibrium boundary conditions on a catalytic surface within the Direct Simulation Monte Carlo (DSMC) method. The radiative equilibrium boundary condition is based on the principle of energy conservation at each surface element, enabling the accurate capture of spatially varying surface temperatures and heat fluxes encountered during atmospheric re-entry. The surface catalycity is represented through the finite-rate surface chemistry (FRSC) model, specifically focusing on the heterogeneous recombination of atomic oxygen on silica surfaces. Both the FRSC model and the radiative equilibrium boundary conditions within the DSMC framework are validated through comparison to analytical solutions. Numerical simulations are conducted for rarefied hypersonic flow around a two-dimensional cylinder under representative re-entry conditions for both non-catalytic and catalytic surfaces. The results demonstrate significant discrepancies in computed surface properties between the radiative equilibrium and conventional isothermal boundary conditions. Furthermore, linear interpolation between results from two independent isothermal boundary conditions is shown to be inadequate for accurately predicting surface heat flux, particularly when surface reactions are considered. The observed discrepancies originate from a non-linear correlation between surface temperature and heat flux, influenced by factors such as surface catalycity and local geometric variations along the cylinder. These findings highlight the necessity of implementing radiative equilibrium boundary conditions within DSMC to ensure physically accurate aerothermodynamic computations.

    \end{abstract}



\end{frontmatter}


\section{Introduction}\label{introduction}  
    During atmospheric re-entry, a high-enthalpy shock layer forms in front of the vehicle forebody, resulting in steep temperature gradients that excite internal energy modes and dissociate gas molecules into atomic species \cite{Leyva2017}. The highly energetic gas particles collide with the vehicle surface, significantly increasing both the incoming heat flux, $q_{in}$, and the surface temperature, $T_s$. These two parameters play a key role in gas–surface interactions during re-entry, influencing phenomena such as hypersonic boundary layer transition \cite{Gai2024} and the thermal response of materials under extreme heating conditions \cite{Lachaud2014, Munafo2022, Padovan2024}. Therefore, accurate predictions of local $q_{in}$ and $T_s$ are essential for the aerothermodynamic analysis of atmospheric re-entry. In doing so, numerical tools such as computational fluid dynamics (CFD) and direct simulation Monte Carlo (DSMC) are employed due to the challenges associated with replicating re-entry conditions experimentally. In particular, the DSMC method has been validated to effectively capture non-equilibrium effects typical of the rarefied hypersonic flow conditions encountered during atmospheric re-entry \cite{Bird1994}.

    Considering $q_{in}$ solely as a result of gas-surface interactions, $q_{in}$ primarily comprises two components: convective heat flux, $q_{conv,in}$, and chemical heat flux, $q_{chem,in}$ \cite{Fay1958}. The former arises from thermal energy exchanges between gas particles and the surface during gas-surface collisions. The latter results from the reaction of gas species on a catalytic surface. The accumulation of $q_{in}$ is prevented through energy dissipation from re-entry vehicle surfaces, as $q_{out}$ \cite{Le2021}. In the absence of external methods of heat dissipation, the thermal radiation heat flux, $q_{rad,out}$, emitted from vehicle surfaces acts as the primary mechanism for heat dissipation \cite{Saad2023, Maout2025}. To satisfy the principle of energy conservation at the gas-surface interface, $q_{in}$ must be in equilibrium with the dissipating heat flux, $q_{out}$.
    
    \begin{figure}[h!]
        \centering
        \includegraphics[width=0.7\textwidth]{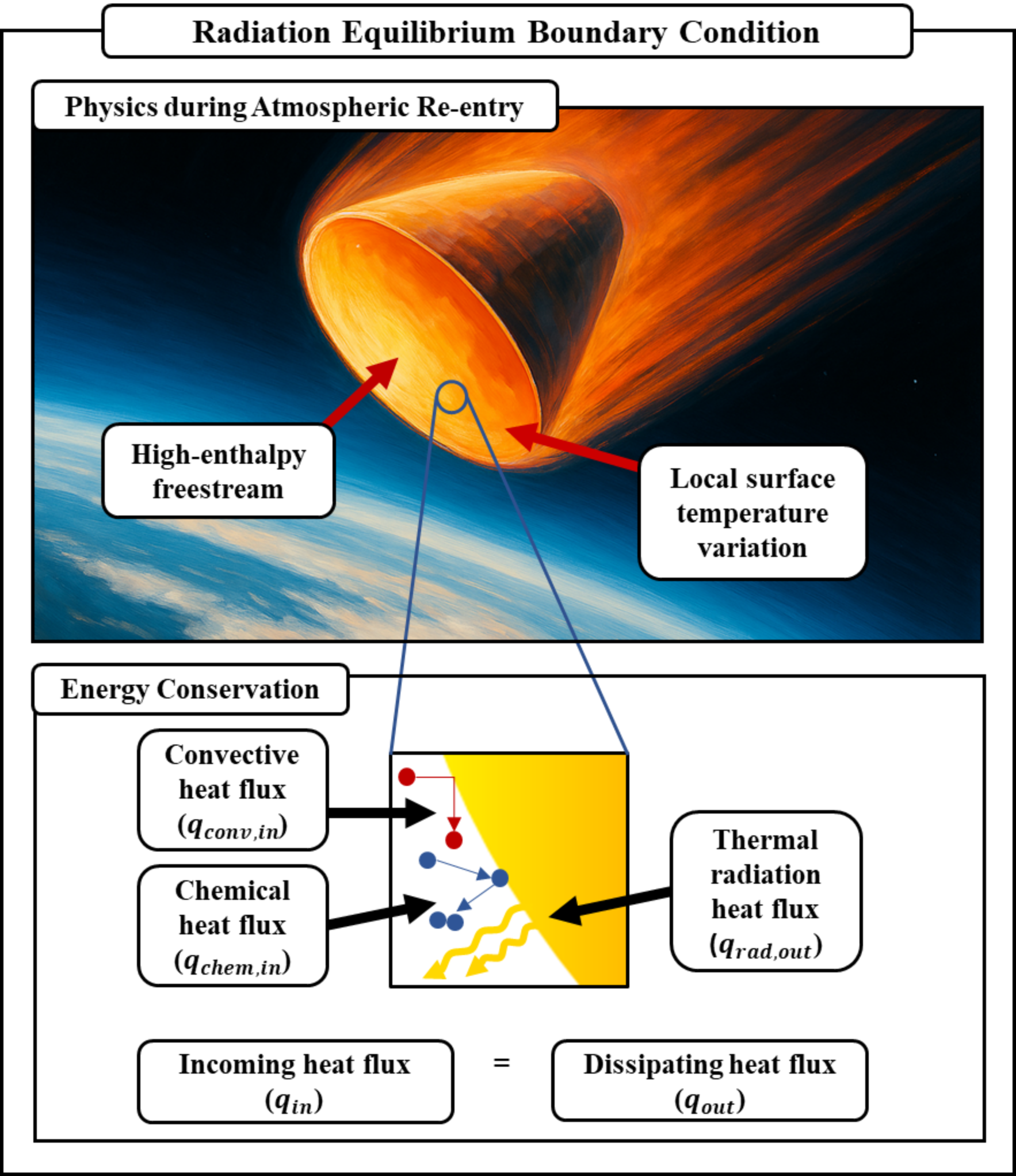}
        \caption{\label{fig:image} Schematic of radiative equilibrium boundary condition on a catalytic surface during atmospheric re-entry.}
    \end{figure}
    
    Numerical investigations using the DSMC method for re-entry aerothermodynamics have traditionally employed an isothermal boundary condition (BC), in which a predetermined and spatially uniform $T_s$ is prescribed across the vehicle geometry \cite{Jun2013, Jun2011, Lofthouse2008}. While computationally convenient, this approach neglects the spatial variability of $T_s$ that naturally arises during atmospheric re-entry. Since $q_{in}$ is dependent on $T_s$, atmospheric re-entry simulations using isothermal BC may not accurately predict the local distribution of $q_{in}$ on the vehicle surface. To enhance accuracy, a common engineering practice is to conduct two independent simulations with isothermal BCs at different $T_s$. The resulting $q_{in}$ values can then be linearly interpolated to estimate the $q_{in}$ at intermediate temperatures. This method relies on the assumption that $T_s$ and $q_{in}$ exhibit a linear correlation across all surface elements. 

    To address the limitations of the isothermal BC, the radiative equilibrium BC can be employed as a more physically consistent alternative. As illustrated in Figure \ref{fig:image}, the radiative equilibrium BC is based on the principle of energy conservation. By enforcing $q_{in} = q_{out}$ at each surface element, the local $T_s$ that would satisfy the energy conservation is computed. This approach makes radiative equilibrium BC particularly appropriate for modeling atmospheric re-entry, where large $T_s$ variations are expected \cite{Maout2025}. The radiative equilibrium BC has been widely adopted in CFD studies of atmospheric re-entry \cite{Wright2006, Pezzella2009, Liu2010, Yang2022, Maout2025}. However, its application within the DSMC framework remains limited and overlooks potential effects of surface reactions \cite{Moss2006, Jin2021}.

    Since the $q_{in}$ and $T_s$ are intrinsically coupled under realistic re-entry conditions, it is important to assess the correlation between these two variables. Several numerical studies have previously investigated the relationship between $T_s$ and $q_{in}$ around the surface of re-entry vehicles. Yang et al. and Luo et al. reported a strong negative linear correlation between $T_s$ and $q_{in}$ near the stagnation point, indicating that an increase in $T_s$ results in a reduction of $q_{in}$ \cite{Yang2022, Luo2020}. Xu et al. attributed the reduction in $q_{in}$ at higher $T_s$ to a decreased thermal gradient between the surface and the energetic freestream, as well as to the reduced compressibility effects at elevated $T_s$ \cite{Xu2021}. However, these studies predominantly assumed non-catalytic surfaces, neglecting realistic surface reactions encountered during actual re-entry conditions. Additionally, they largely focused on forebody regions where high aerothermodynamic load from atmospheric re-entry is expected. Nevertheless, $q_{in}$ at the aftbody remains non-negligible, and surface temperatures there can approach 1000 K along typical re-entry trajectories \cite{Wright2006}. Wright et al. performed CFD simulations of the aftbody of the Apollo re-entry capsule and have shown evidence that high $T_s$ may actually increase the computed $q_{in}$ \cite{Wright2006}. Therefore, resolving the relationship between $T_s$ and $q_{in}$ across the entire vehicle surface, including the aftbody region, with the effect of surface reaction is essential.
    
    The present study aims to employ the radiative equilibrium BC within the DSMC framework that considers spatially varying $T_s$ across the surface for accurate aerothermodynamics analysis of atmospheric re-entry. Specifically, it seeks to: 
    (1) implement and validate the radiative equilibrium BC in DSMC; (2) investigate how the radiative equilibrium BC computes $q_{in}$ and $T_s$ compared to the conventional isothermal BC; and (3) analyze the correlation between $T_s$ and $q_{in}$ under radiative equilibrium BC at different surface catalycity. The remainder of this paper is structured as follows. Section \ref{physicalmodel} introduces the physical models that govern non-equilibrium re-entry flows. Section \ref{numericalmethodology} details the numerical methodology, describing the DSMC with the implementation of the radiative equilibrium BC. Section \ref{validation} presents validation of models implemented in this study, and Section \ref{resultsanddiscussion} discusses the effects of the radiative equilibrium BC on different surface parameters and their interdependent correlation in the atmospheric re-entry DSMC simulation. 

\clearpage
\section{Physical models}\label{physicalmodel}
\subsection{Boltzmann equation}
    The Navier-Stokes equations, which form the foundation of conventional CFD, are based on the continuum assumption, where local thermodynamic equilibrium is maintained. However, these equations become inadequate in effectively resolving atmospheric re-entry, where non-equilibrium effects dominate \cite{Jun2018, Kim2023}. As a more suitable alternative, the Boltzmann equation provides a kinetic description of gas dynamics by describing the evolution of the velocity distribution function $f$ in phase space spanned by position, $\overrightarrow{r}$, and velocity, $\overrightarrow{c}$ \cite{Bird1994, Anderson2019}. This equation takes the following form:
    
    \begin{equation}\label{eqn:boltz}
        \frac{\partial}{\partial t}\left( nf \right) + \overrightarrow{c} \cdot \frac{\partial}{\partial \overrightarrow{r}}\left( nf \right) + \overrightarrow{F} \cdot \frac{\partial}{\partial \overrightarrow{c}}\left( nf \right) = \int_{-\infty}^{\infty}\int_{0}^{4\pi}n^2\left[ f^*f^*_1-ff_1 \right]c_r\sigma_{cs} d \Omega d\overrightarrow{c_1}.
    \end{equation}
    
    \noindent The left-hand side of the equation accounts for particle transport and external forces, $\overrightarrow{F}$, while the right-hand side represents modifications to $f$ due to binary collisions between gas molecules. The relative velocity of the colliding particles, $c_r$, the solid angle of deflection, $\Omega$, and the collision cross-section, $\sigma_{cs}$, dictate the nature of these interactions. The superscript $*$ and the subscript $1$ denote the properties of the post-collision particle and those of the collision partner, respectively.

\subsection{Surface heat flux}
    The rate of thermal energy transfer from the flowfield to the surface per unit area is quantified by the incoming heat flux, $q_{in}$. For typical atmospheric re-entry conditions, this flux is primarily attributed to gas-surface interactions and consists of two dominant components: the convective heat flux, $q_{conv,in}$, and the chemical heat flux, $q_{chem,in}$. In addition to these, radiation heat flux from the flowfield, $q_{rad,in}$, can arise due to the presence of electronically excited gas particles \cite{Bird1987}. However, for entry velocities below approximately 10 km/s, the contribution of $q_{rad,in}$ to over $q_{in}$ is negligible \cite{Goulard1961, Park2004}. Therefore, this study assumes that $q_{in}$ results solely from gas-surface interactions, such that $q_{in} = q_{conv,in} + q_{chem,in}$. 
    
    When $T_s$ rises as a result of $q{in}$, the heated surface naturally dissipates heat at a rate per unit area defined by the outgoing heat flux, $q_{out}$. The primary dissipation mechanism is thermal re-radiation to the surroundings, which is quantified as $q_{rad,out}$ \cite{Maout2025}. For certain materials, additional heat removal may occur via material response processes such as ablation or pyrolysis, represented by $q_{mat,out}$ \cite{Johnston2013}. However, such effects are not considered in this study. In addition, conductive heat flux, $q_{cond,out}$, can dissipate heat from hot surfaces to adjacent surface or bulk elements. Assuming that internal heat conduction is negligible within the vehicle \cite{Saad2023}, heat dissipation on surface elements can only occur through thermal radiation, such that $q_{out}=q_{rad,out}$. Under steady-state conditions, energy conservation at the gas-surface interface requires that the $q_{in} = q_{out}$. This balance can be expressed as: 
    
    \begin{equation}\label{eqn:qbalance} 
        q_{conv,in} + q_{chem,in} = q_{rad,out}.
    \end{equation}

    \noindent The heat flux components considered in this study are discussed in more detail.

    \subsubsection{Convective heat flux}
    The $q_{conv,in}$ is the result of the direct energy exchange between gas particles and the surface. According to the principle of energy conservation, the energy lost by the gas particles is equivalent to the energy gained by the surface during each gas-surface collision. Hence, the energy transfer to the surface can be inferred from the net difference in the energy carried by gas particles before and after their collision with the surface. Consequently, $q_{conv,in}$ is determined as the total energy lost by the gas particles per unit area, $A_s$, and per unit time, $\Delta t$, across all gas-surface collisions such that,

    \begin{equation}\label{eqn:qconv} 
        q_{conv,in} = \frac{\sum \left({E}_{in}-{E}_{out} \right)}{A_s \Delta t},
    \end{equation}

    \noindent where $E_{in}$ and $E_{out}$ represent the energy of the gas particles before and after their collision with the surface, respectively. 
    
    Different gas-surface interaction models have been developed to approximate the ${E}_{out}$ distributions of particles \cite{Cercignani1971, Huh2025}. Among these models, the Maxwell model is commonly utilized due to its simplicity \cite{Maxwell1878}. The Maxwell model incorporates the thermal accommodation coefficient, $\alpha$, to quantify the extent to which gas particles achieve thermal equilibrium with the surface upon impact. This coefficient is defined as follows,

    \begin{equation}\label{eqn:TAC} 
        \alpha = \frac{\bar{E}_{in}-\bar{E}_{out}}{\bar{E}_{in}-\bar{E}_{s}(T_s)}, 
    \end{equation}

    \noindent where $\bar{E}_{in}$ and $\bar E_{out}$ denote the average energies of incoming and outgoing gas particles, respectively, while $\bar E_s(T_s)$ represents the Maxwell-Boltzmann energy distribution corresponding to $T_s$. In other words, $\alpha=1$ represents the diffuse reflection of particles, where the post-collision energy is a function of $T_s$ only, and $\bar E_{out} = \bar E_s(T_s)$ Conversely, $\alpha = 0$ results in specular reflections of particles, where $\bar{E}_{in} = \bar{E}_{out}$, irrespective of $T_s$. Due to surface corrugation and contamination typically encountered by vehicle surface during atmospheric re-entry, the $\alpha$ is commonly approximated to $\alpha \approx 1$ \cite{Huh2025, Park2025}.
    
    \subsubsection{Chemical heat flux}
    The chemical heat flux, $q_{chem,in}$, originates from exothermic reactions occurring at the surface. The dominant surface reaction during re-entry is the heterogeneous recombination of $O$, which can account for more than 50\% of $q_{in}$ during atmospheric re-entry. \cite{Molchanova2018, Candler2019}. This process involves the adsorption of $O$ atoms onto active sites, followed by their recombination, which can be categorized into two distinct mechanisms: the Eley-Rideal (ER) mechanism and the Langmuir-Hinshelwood (LH) mechanism. In the ER mechanism, an incoming gas-phase atom directly reacts with an adsorbed atom, resulting in molecule formation upon impact. Conversely, the LH mechanism involves two gas-phase atoms adsorbed at the surface, which can diffuse across the surface and recombine. The rates of the two recombination mechanisms dictate the release of chemical energy, $\Delta E_{chem}$, which can be translated into $q_{chem,in}$ at the surface elements as:

    \begin{equation}\label{eqn:qchem} 
        q_{chem,in} = \frac{\sum \Delta E_{chem}}{A_s \Delta t}
    \end{equation}

    \noindent with the assumption that $\Delta E_{chem}$ is fully accommodated into the surface \cite{Ko2024}.
        
    Surface reactions are fundamentally governed by the principles of chemical kinetics that define the rates at which reactions occur. In numerical simulations, this physical phenomenon is represented using the finite-rate surface chemistry (FRSC) model \cite{Sorensen2013, Molchanova2018, Swaminathan-Gopalan2024}. The FRSC model employs temperature-dependent reaction rate coefficient, $k$, expressed using the Arrhenius equation for each potential surface reaction, such that, 
    
        \begin{equation}\label{eqn:arrhenius}
            k(T_s) = A T_s^{b} exp(-E_{a}/{k_BT_s}), 
        \end{equation}
    
    \noindent where $k_B$ denotes the Boltzmann constant. The parameters $A$, $b$, and $E_a$ correspond to the pre-exponential factor, the temperature exponent, and the activation energy, respectively. The explicit dependence of $k$ on $T_s$ emphasizes the critical role of $T_s$ in determining surface catalycity and the value of $q_{chem,in}$.
    
    \subsubsection{Thermal radiation heat flux}
    Thermal radiation heat flux, $q_{rad,out}$, refers to the rate of energy dissipation per unit area from heated surfaces in the form of photons or electromagnetic waves. The $q_{rad,out}$ results from atomic and molecular oscillations driven by internal energy of the surface matter, which is directly related to the $T_s$ \cite{Incropera2013}.
    
    The emitted $q_{rad,out}$ is governed by the Stefan-Boltzmann law, mathematically expressed as,
    
    \begin{equation}\label{eqn:radiation1}
        q_{rad,out} = \epsilon \sigma T_s^4,
    \end{equation}

    \noindent where $\epsilon$ represents the emissivity of the surface, and $\sigma$ denotes the Stefan-Boltzmann constant. The $\epsilon$ is a material property that quantifies the effectiveness of thermal radiation emission. Its value ranges between $0 \leq \epsilon \leq 1$, with higher values indicating greater emissive efficiency.     
\subsection{Surface boundary conditions}
    Due to its simplicity, isothermal BC has been conventionally employed for numerical studies of re-entry flows, in which each surface element is assumed to maintain a uniform $T_s$, expressed as,

    \begin{equation}\label{eqn:isothermal_T} 
    T_{s,i} = constant,
    \end{equation}

    \noindent where $i$ denotes the $i^{th}$ surface element. However, realistic flight conditions exhibit spatial variations in $T_s$ across the vehicle surface due to differences in aerodynamic heating, which are associated with their relative positions with respect to the freestream \cite{Wright2006, Maout2025}. Consequently, the isothermal BC inherently fails to accurately represent the realistic $T_s$. Moreover, $q_{conv,in}$ and $q_{chem,in}$ are found to be functions of local $T_s$. Therefore, the application of an isothermal BC can undermine the accuracy of $q_{in}$ computations using numerical methods.

    Since all considered heat flux components are functions of $T_s$, it is possible to determine the value of $T_s$ that would satisfy the principles of energy conservation at the gas-surface interface outlined in Equation \ref{eqn:qbalance}. By enforcing this energy balance at each surface element, the radiative equilibrium BC can model spatially varying $T_s$ as below:
    
    \begin{equation}\label{eqn:rebc_T} 
        T_{s,i} = \sqrt[4]{\frac{q_{in,i}(T_{s,i})}{\epsilon \sigma}}.
    \end{equation}
    
    \noindent Therefore, radiative equilibrium BC is a more physically consistent alternative to isothermal BC. This approach inherently accounts for local thermodynamic conditions and can provide improved accuracy in $q_{in}$ predictions compared to the isothermal BC. Note that $T_{s,i}$ and $q_{in,i}$ are interdependent variables, which makes Equation \ref{eqn:rebc_T} implicit. Therefore, solving for this equation at each $i^{th}$ surface for the radiative equilibrium BC requires an iterative numerical method.

\clearpage
\section{Numerical methodology}\label{numericalmethodology}
    \subsection{Direct simulation Monte Carlo}
\begin{figure}[h!]
        \centering
        \includegraphics[width=0.95\textwidth]{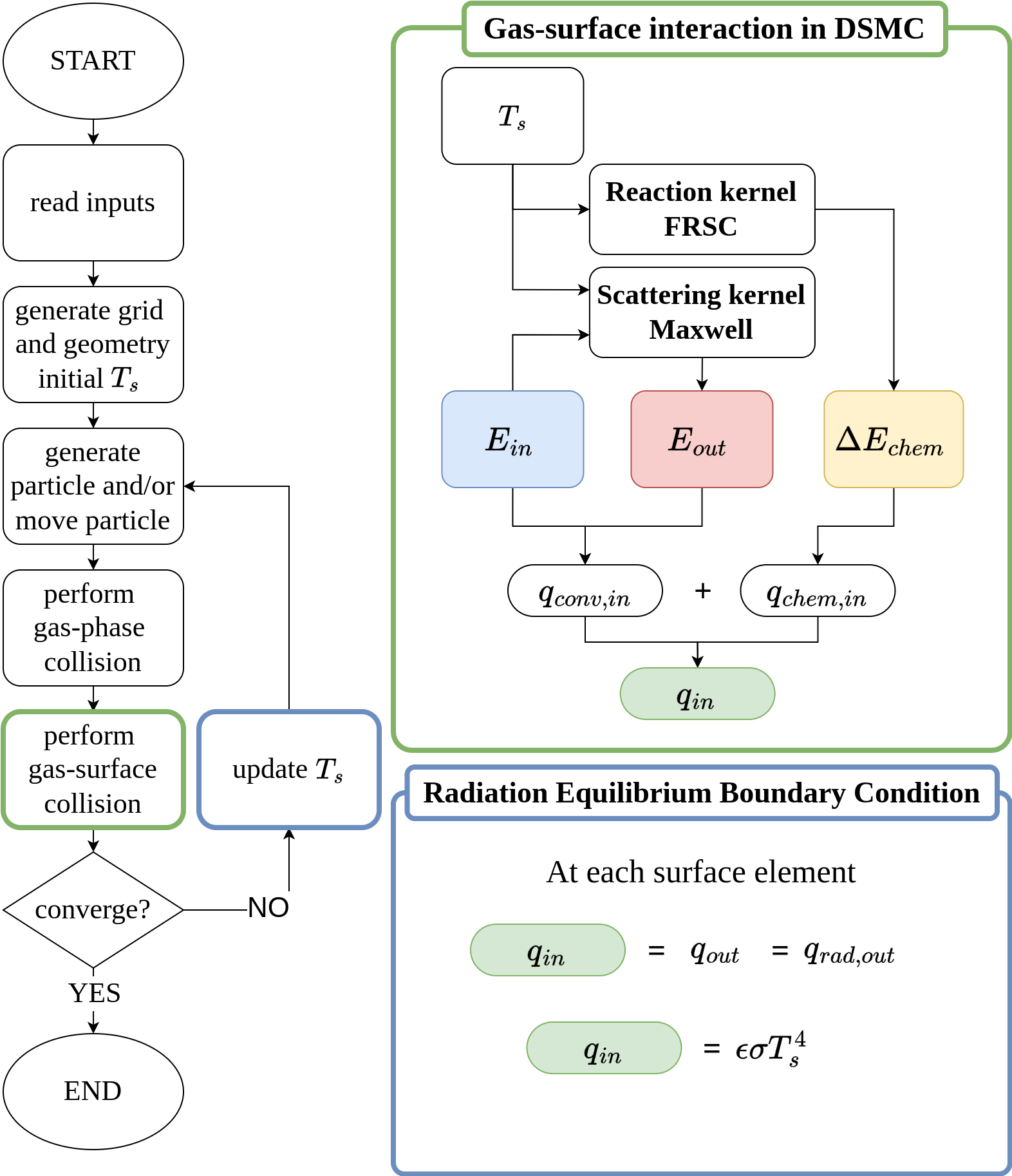}
        \caption{\label{fig:method_rebc} Flowchart of the DSMC method implementing the radiative equilibrium boundary condition coupled with the FRSC model.}
    \end{figure}

     Although the Boltzmann equation provides an accurate framework for modeling non-equilibrium gas dynamics, obtaining its analytical solution remains challenging due to the complexity of the collision integral. Therefore, numerical approaches, particularly the Direct Simulation Monte Carlo (DSMC) method, have been widely employed \cite{Bird1994, Bird2013, Jun2018b, Kim2022, Kim2024, Kim2024a, Park2024}. This study utilizes the Stochastic PArallel Rarefied-gas Time-accurate Analyzer (SPARTA), an open-source DSMC solver developed by Sandia National Laboratories \cite{Gallis2014}. The schematic algorithm of the DSMC method is illustrated in Figure \ref{fig:method_rebc}. The DSMC begins with the setup of the computational grid and geometry. Then, it proceeds with the generation of simulation particles that represent a collection of real gas particles. At each time step, the particles are made to move within or across grid cells. During this process, particles may undergo gas-gas collisions, modeled stochastically to approximate the Boltzmann equation \ref{eqn:boltz}, or interact with surfaces through gas-surface collisions that are defined by the prescribed surface geometry.

\clearpage

\clearpage

     The gas-surafce interaction is detailed in the green box of Figure \ref{fig:method_rebc}. When gas particles with energy $E_{in}$ collide with a surface element at temperature $T_s$, the scattering kernel determines $E_{out}$, while the reaction kernel evaluates $\Delta E_{chem}$. In this study, the Maxwell model is used as the scattering kernel, and the FRSC model is employed for the reaction kernel. Consequently, the $q_{conv,in}$ is computed using Equation \ref{eqn:qconv} and $q_{chem,in}$ is computed using \ref{eqn:qchem}. The sum of the two heat flux components can give $q_{in}$ of each surface element.
     
    The FRSC model implemented as a reaction kernel in the DSMC code SPARTA was developed by Swaminathan-Gopalan et al. \cite{Swaminathan-Gopalan2024}. It converts the active site density, $n_s$, and $k$ of each possible surface reaction into the reaction probability, $P_{react}$. Upon each gas-surface collision, the surface reaction is stochastically evaluated by either accepting or rejecting the surface reaction with a comparison of $P_{react}$ against a generated random number. Upon acceptance, the gas species transforms into a different chemical species, effectively simulating the surface reaction. When utilizing the FRSC module within the DSMC solver SPARTA, special attention must be paid to the input values of $k$. In the majority of the literature, Arrhenius-based $k$ is expressed using conventional gas-phase kinetics \cite{Norman2012, Deutschmann1995}. However, the FRSC module in SPARTA uses the dimensionless reaction rate coefficient \cite{Swaminathan-Gopalan2024}. Accordingly, $k$ values used in the study with physical units are non-dimensionalized to ensure consistency in modeling surface reactions.

    \subsection{Radiative equilibrium boundary condition}
    The blue box of Figure \ref{fig:method_rebc} highlights how radiative equlibirum BC is implemented in DSMC. The principle of energy conservation is enforced at each surface element and timestep by iteratively updating $T_s$ based on the calculated $q_{in}$ using Equation \ref{eqn:rebc_T}. Then, it is utilized for subsequent iterations of the DSMC algorithm, particularly for the gas-surface interaction kernels in the green box of Figure \ref{fig:method_rebc}. To account for the radiative equilibrium BC on catalytic surfaces, modifications have been implemented in SPARTA DSMC to integrate with the scheme for updating $T_s$.

\clearpage
\section{Validation of models}\label{validation}
\subsection{Finite-rate surface chemistry}\label{frsc}
    The catalytic condition considered in this study is the heterogeneous recombination of $O$ on the silica ($SiO_2$) surface. The associated reaction sets were identified by Norman et al. through ab-initio molecular dynamics simulations \cite{Norman2011, Norman2012a}. The objective of this validation phase is to apply the FRSC model within the DSMC framework using the given reaction sets, evaluate the resulting surface recombination coefficient, $\gamma$, and compare it against analytical solutions and experimental data. The $\gamma$ characterizes the catalytic efficiency of a surface by representing the fraction of incident atomic species that recombine upon collision \cite{Candler2019}. It is defined as,

    \begin{equation}\label{eqn:gamma} 
        \gamma = \frac{j_{O,recomb}}{j_{O,in}},
    \end{equation}

    \noindent where $j_{O,in}$ represents the flux of $O$ species to the surface, and $j_{O,recomb}$ denotes the flux of $O$ that undergo recombination. The analytical calculations used for comparison are obtained by solving the systems of reaction rate equations, assuming steady-state at the gas-surface interface. Experimental values of $\gamma$ are derived from independent studies of the $O$ reaction on $SiO_2$ surfaces \cite{Paul2023}. In these experiments, neutral atomic oxygen was typically generated using low-pressure microwave discharges \cite{Balat-Pichelina2003, Kim1991, Marschall1997}, with the exception of the study by Stewart et al., who employed an arc-jet DC discharge \cite{Stewart1997}. The DSMC simulations are set up to replicate the experimental test conditions, in which a flat $SiO_2$ plate is exposed to $O$ of 100 Pa. The range of $T_s$ from 400 K to 2000 K is used, which falls within the expected $T_s$ range for the re-entry vehicle surface \cite{Wright2006}. The active sites density, $n_s$, represents the number of adsorption sites available for $O$ on the $SiO_2$ surface. Physically, $n_s$ depends on the surface morphology and corrugation, which are challenging to accurately characterize experimentally \cite{Huh2025}. Norman et al. note that arbitrary choice of $n_s$, without loss of generality, does not alter the general trend of $\gamma$ as a function of $T_s$, but instead affects its overall magnitude \cite{Norman2011, Norman2012a}. Thus, following the approach of Norman et al., this study treats $n_s$ as an adjustable parameter to align model predictions with experimental results.

\begin{figure}[h!]
        \centering
        \subfigure[High-T model. $n_s = 10^{18}$ m$^{-2}$.]{
            \includegraphics[width=0.45\textwidth]{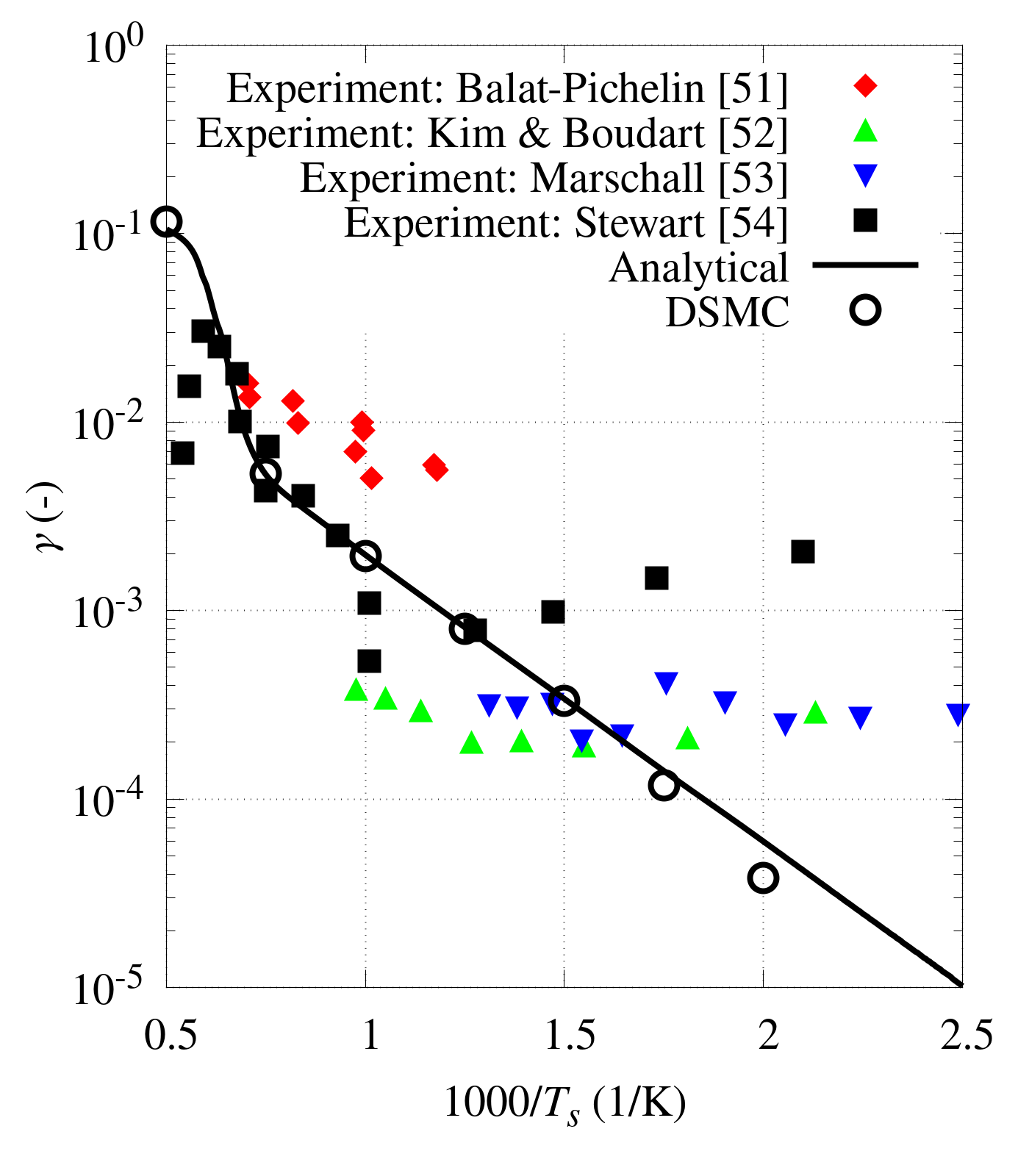}
            \label{fig:gamma_high18}
        }
        \subfigure[Low-T model. $n_s = 10^{17}$ m$^{-2}$.]{
            \includegraphics[width=0.45\textwidth]{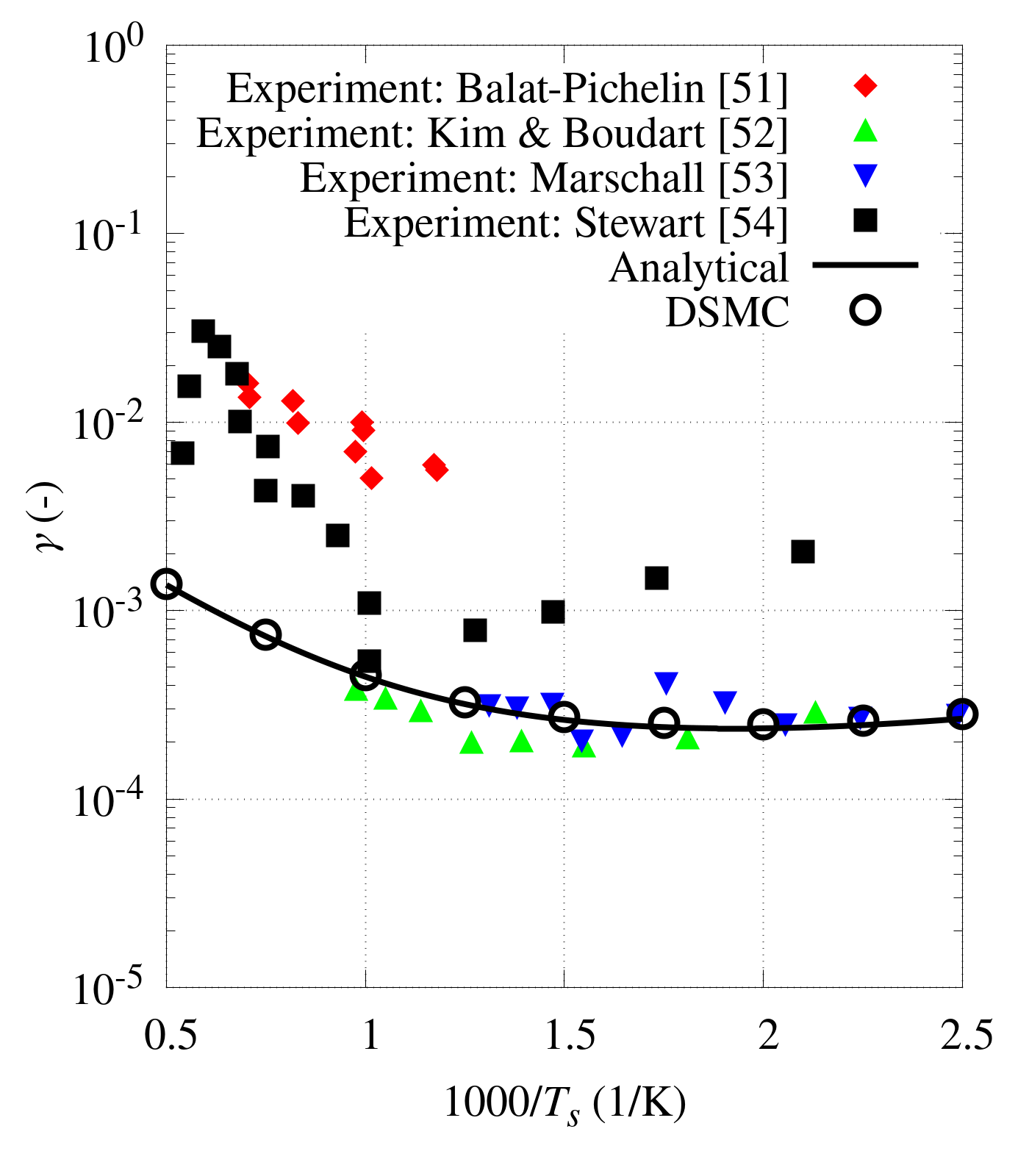}
            \label{fig:gamma_low17}
        }
        \subfigure[All-T model. $n_s = 3 \times 10^{17}$ m$^{-2}$.]{
            \includegraphics[width=0.45\textwidth]{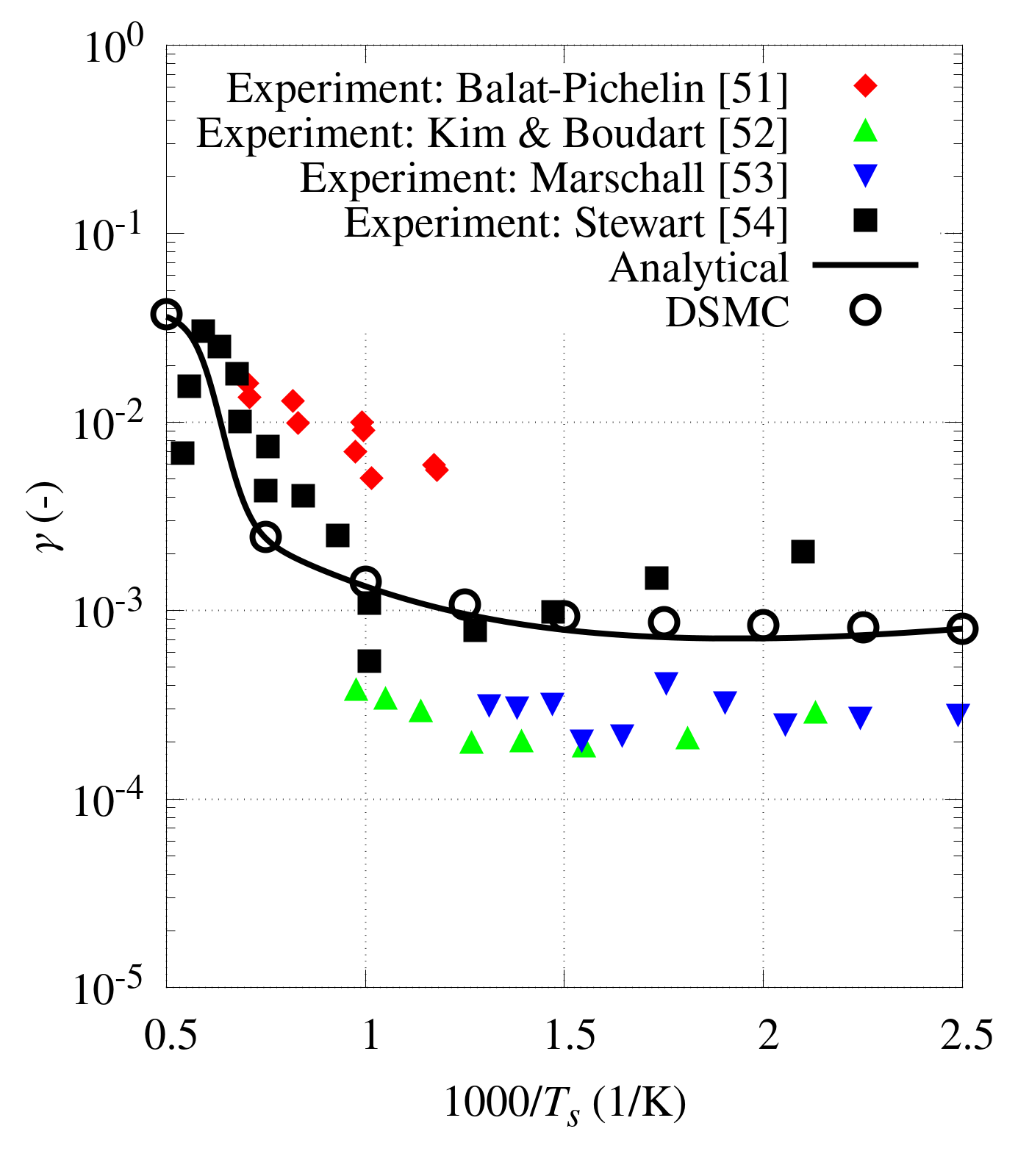}
            \label{fig:gamma_all317}
        }
        \caption{$\gamma$ as a function of $1000 / T_s$.}
        \label{fig:gamma}
    \end{figure}

    Figure \ref{fig:gamma} shows $\gamma$ as a function of the reciprocal of $T_s$. Solid symbols, solid lines, and open symbols represent the experimental, analytical, and DSMC simulation results, respectively. Initially, two distinct recombination models proposed by Norman et al. are considered: a high-temperature (high-T) model based on the ER mechanism and a low-temperature (low-T) model based on the LH mechanism \cite{Norman2012a}. Figure \ref{fig:gamma_high18} presents $\gamma$ for the high-T model, which only considers ER recombination, at $n_s = 10^{18}$ m$^{-2}$. At high $T_s$ ($T_s >$ 1000 K), experimental results show an exponential increase in $\gamma$ with increasing $T_s$. Physically, this trend can be explained by the nature of the ER recombination mechanism. The ER recombination occurs between an incident gaseous $O$ and an adsorbed $O$ on the surface active site. Since adsorbed $O$ atoms achieve thermal equilibrium with the surface \cite{Hammond2013}, they will attain high internal energy levels when adsorbed to surfaces with high $T_s$. Consequently, collisions between gaseous $O$ atoms and the energetically excited adsorbed species at high $T_s$ are more likely to exceed the $E_a$ barrier required for recombination. This exponential behavior is well-reproduced by the analytical results of the high-T model, indicating that ER recombination dominates on $SiO_2$ surfaces at high $T_s$. Additionally, the DSMC simulations align closely with the analytical predictions across the entire $T_s$ range examined, validating the accurate implementation of the FRSC model within DSMC. However, at lower $T_s$ ($T_s <$ 1000 K), experimental data reveal a relatively constant $\gamma$ with $T_s$. This trend is not captured by either the analytical or DSMC results of the high-T model. The discrepancy arises due to the neglect of the LH mechanism, which is not incorporated in the high-T model \cite{Norman2012a}. Figure \ref{fig:gamma_low17} illustrates the results of the low-T model, which exclusively considers the LH recombination mechanism, at $n_s = 10^{17}$ m$^{-2}$. The relatively constant $\gamma$ at low $T_s$ ($T_s <$ 1000 K) can be physically explained by the LH mechanism. In the LH recombination process, two adsorbed oxygen atoms diffuse across the surface until they recombine. At higher $T_s$, adsorbed atoms have a higher rate of desorption before diffusion may occur, which restricts the exponential increase in $\gamma$ seen for the ER mechanism. In contrast, at low $T_s$, adsorbed atoms remain on the surface longer, increasing their probability of diffusing and recombining \cite{Norman2012a}, to an extent where it offsets the exponential decrease of $\gamma$. This phenomenon accounts for the plateau of $\gamma$ as a function of $T_s$. Both analytical and DSMC computations closely replicate this trend and align with experimental data, particularly at lower $T_s$. Thus, these findings suggest that the LH recombination mechanism dominates the heterogeneous recombination process at lower $T_s$. However, the model does not capture the exponential increase of $\gamma$ observed at higher $T_s$ due to the exclusion of the ER recombination mechanism.

    To accurately capture the $\gamma$ within the range of $T_s$ considered, an all-temperature (all-T) model has been developed by combining the reaction mechanisms used in both the high-T and low-T models. Figure \ref{fig:gamma_all317} illustrates the resulting $\gamma$ from the all-T model at an active site density of $n_s = 3 \times 10^{17}$ m$^{-2}$. Both analytical calculations and DSMC simulations employing the all-T model effectively replicate the experimental trends of $\gamma$ at both high and low $T_s$. Specifically, at high $T_s$, $\gamma$ exhibits an exponential decrease with decreasing $T_s$ due to the dominance of the ER recombination mechanism. Conversely, at low $T_s$, the $\gamma$ remains relatively constant as $T_s$ decreases, attributed to the dominance of the LH recombination mechanism. By incorporating both ER and LH recombination processes, the all-T model successfully captures the characteristics of $\gamma$ dependent on $T_s$ observed experimentally. Since this study aims to investigate the effects of spatial variations in $T_s$ along re-entry vehicle surfaces through numerical DSMC simulations, the use of the all-T model ensures an accurate representation of surface catalytic behavior on $SiO_2$ across a broad $T_s$ range expected during atmospheric re-entry.

\clearpage

\clearpage
\subsection{Radiative equilibrium boundary condition}
    Next, the implementation of the radiative equilibrium BC within the DSMC framework is validated. DSMC simulations are conducted using conditions representative of a specific point along the Apollo AS-202 re-entry trajectory, specifically at 4750 seconds after launch \cite{Wright2006}. The freestream conditions employed in these simulations are summarized in Table \ref{tab:freestream}, with species-specific number densities, $n_\infty$, derived using the NRLMSIS 2.0 atmospheric model \cite{Emmert2021}. The computational geometry is a standard benchmark configuration consisting of a 2D cylinder with a radius of 0.1524 m \cite{Lofthouse2008}. In simulating gas-surface interactions, complete thermal accommodation ($\alpha = 1$) is assumed. Since the radiative equilibrium BC in DSMC should be consistent irrespective of the surface conditions, both non-catalytic and catalytic surfaces are validated. The all-T model of $SiO_2$ is used to simulate catalytic surfaces. An active site density of $n_s = 10^{18}$ m$^{-2}$ is used to reflect realistic surface conditions influenced by corrugation and erosion during atmospheric re-entry \cite{Yang2019a}. Additionally, it is assumed that the chemical energy released from exothermic recombination reactions is fully accommodated by the surface \cite{Ko2024}. To further examine the sensitivity of the radiative equilibrium BC to surface emissivity, simulations are performed with emissivity values of $\epsilon = 0.3$ and $\epsilon = 0.9$.

    \begin{table}[h!]
        \centering
        \begin{tabular}{ccccccc}
            \toprule
            Time  & Altitude  & $Ma_{\infty}$ & $V_{\infty}$  & $T_{\infty}$  & $n_{N_2,\infty}$ & $n_{O_2,\infty}$ \\
            (s) & (km) & (m/s) & (K) & ($\times 10^{20}$ 1/m$^3$) & ($\times 10^{20}$ 1/m$^3$) \\
            \hline
            4750   & 74.5  & 22.0 & 6390   & 210 & 7.416 &  1.984   \\
            \bottomrule
        \end{tabular}
        \caption{Freestream conditions.}
        \label{tab:freestream}
    \end{table}

\begin{figure}[h!]
        \centering
        \includegraphics[width=0.7\textwidth]{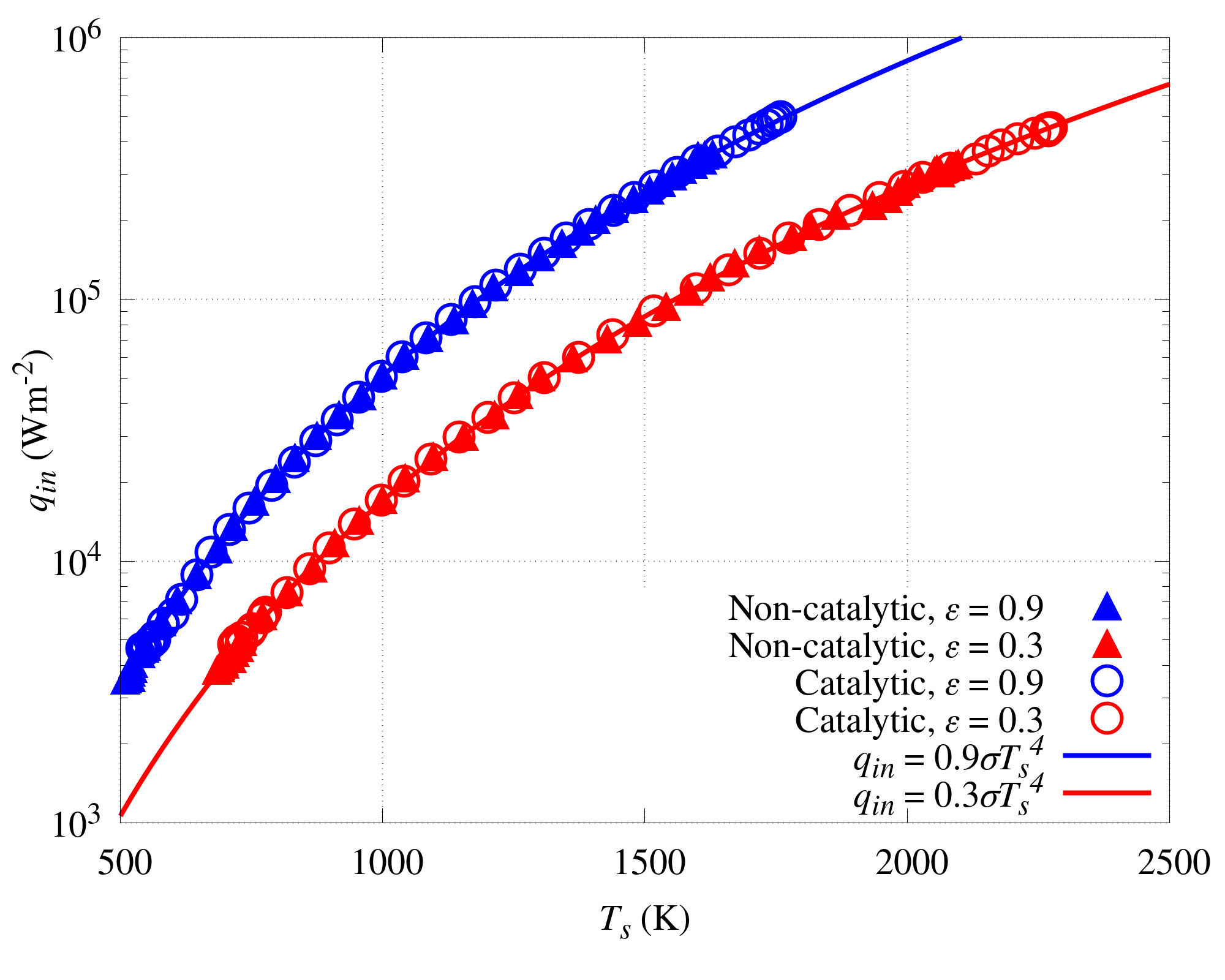}
        \caption{\label{fig:valid_rebc}$q_{in}$ as a function of $T_s$.}
    \end{figure}

    Figure \ref{fig:valid_rebc} presents the $q_{in}$ as a function of $T_s$ for different $\epsilon$ and surface catalycity. Symbols represent each surface element with varying local $T_s$ and $q_{in}$ computed along the 2D cylinder using radiative equilibrium BC in DSMC. The solid lines depict analytical solutions following the Stefan–Boltzmann law as Equation \ref{eqn:radiation1}. Symbol colors distinguish between $\epsilon=0.9$ (blue) and $\epsilon=0.3$ (red) cases, while symbol types differentiate between non-catalytic (triangles) and catalytic (circles) surface conditions. The DSMC results closely align with the analytical solutions across all surface elements, confirming the successful validation of the implemented radiative equilibrium BC. Furthermore, at a given $q_{in}$, surfaces with higher $\epsilon = 0.9$ (in blue) consistently exhibit lower $T_s$ compared to those with lower $\epsilon = 0.3$ (in red). This physically occurs because higher $\epsilon$ can promote efficient thermal radiation emission from the surface, facilitating effective heat dissipation and preventing excessive rise of $T_s$. Additionally, the catalytic surfaces exhibit a range of $q_{in}$ and $T_s$ that are higher than those of non-catalytic counterparts at the same $\epsilon$ due to exothermic surface reactions. However, the inclusion of $q_{chem,in}$ should not alter the energy conservation at the gas-surface interface described by Equation \ref{eqn:qbalance}. The implemented radiative equilibrium BC in DSMC shows the expected behavior as catalytic and non-catalytic surfaces converge onto a single trend line following the energy conservation equation.

\clearpage
    
\clearpage

\section{Surface properties using radiative equilibrium boundary condition}\label{resultsanddiscussion}
    \subsection{Comparison to isothermal boundary condition}
    A series of DSMC simulations is conducted using a 2D cylinder with a radius of 0.1524 m under atmospheric re-entry conditions. The freestream conditions used in all simulations are summarized in Table \ref{tab:freestream}. The radiative equilibrium BC is applied and compared against isothermal BCs. Four cases are considered, depending on the selection of BCs as outlined in Table \ref{tab:cases}. Cases A and B employ isothermal BCs with fixed, uniform $T_s$ of 1300 K and 2000 K, respectively. These temperatures fall within the expected $T_s$ range for the Apollo capsule during re-entry \cite{Park2004}. Case C uses the radiative equilibrium BC. The value of $\epsilon$ is assumed to be $0.9$, as materials within the range of $0.8 \leq \epsilon \leq 1$ are typically selected for re-entry vehicles \cite{He2009}. Case D implements linear interpolation between Cases A and B to reconstruct the $q_{in}$ corresponding to the $T_s$ profile from Case C. Each case is simulated with both non-catalytic and catalytic surface conditions. Surface catalycity is modeled using the FRSC model within DSMC, which accounts for the heterogeneous recombination of $O$ on $SiO_2$ with $n_s = 10^{18}$ m$^{-2}$.

    \begin{table}[h!]
        \centering
        \begin{tabular}{ccc}
            \toprule
                Case & Boundary Condition & $T_s$  \\
            \hline
                A & Isothermal & 1300 K  \\
                B & Isothermal & 2000 K  \\
                C & Radiative equilibrium & locally variable  \\
                D & Isothermal: Linear interpolation & locally variable$^{*}$  \\
            \bottomrule
            \multicolumn{3}{l}{\small * matches local $T_s$ distribution of Case C.} \\

        \end{tabular}
        \caption{Surface conditions for each case.}
        \label{tab:cases}
    \end{table}

\begin{figure}[h!]
        \centering
        \subfigure[$q_{in}$ as a function of $\theta$.]{
            \includegraphics[width=0.475\textwidth]{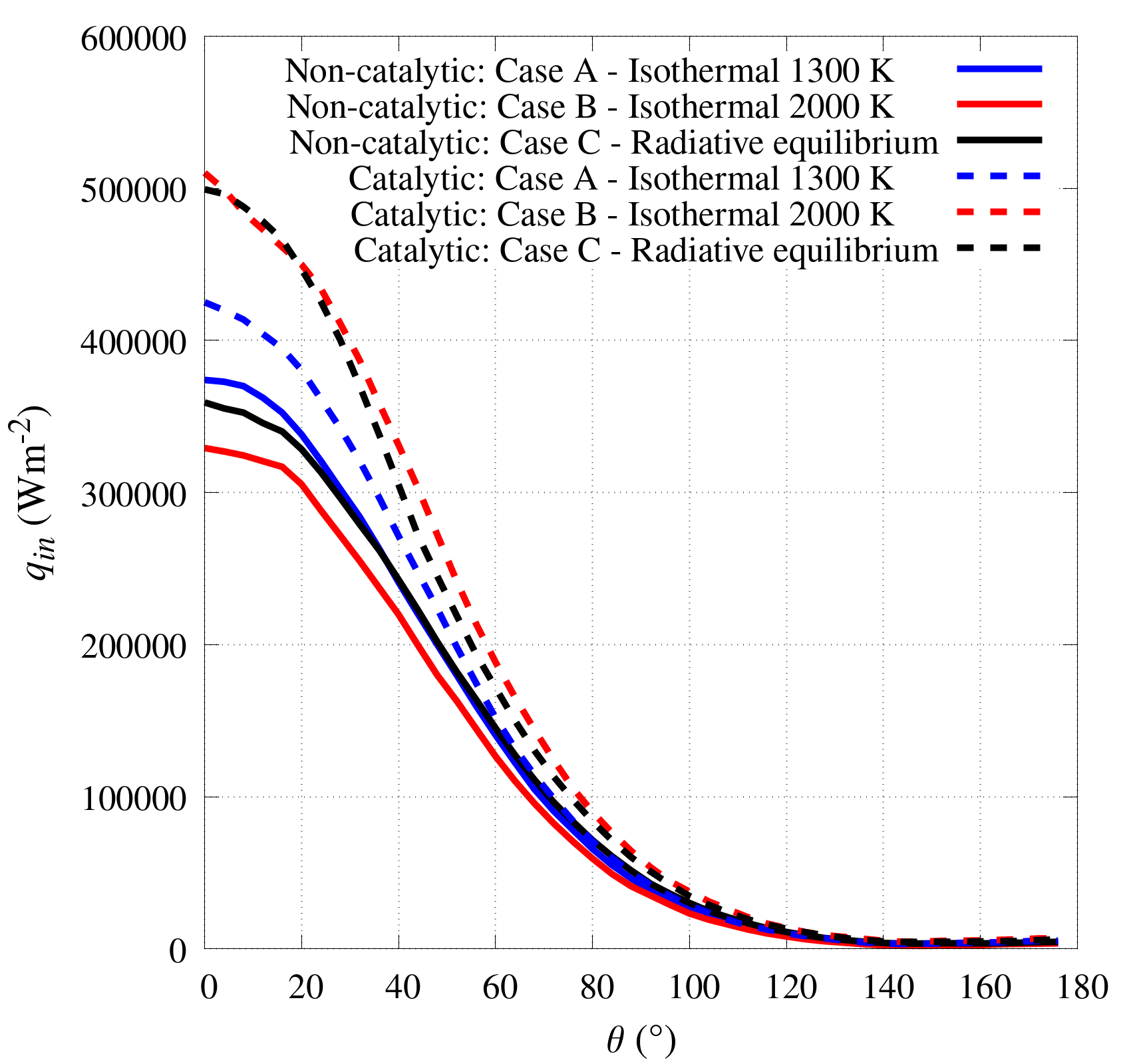}
            \label{fig:heatflux2}
        }
        \subfigure[$T_s$ as a function of $\theta$.]{
            \includegraphics[width=0.475\textwidth]{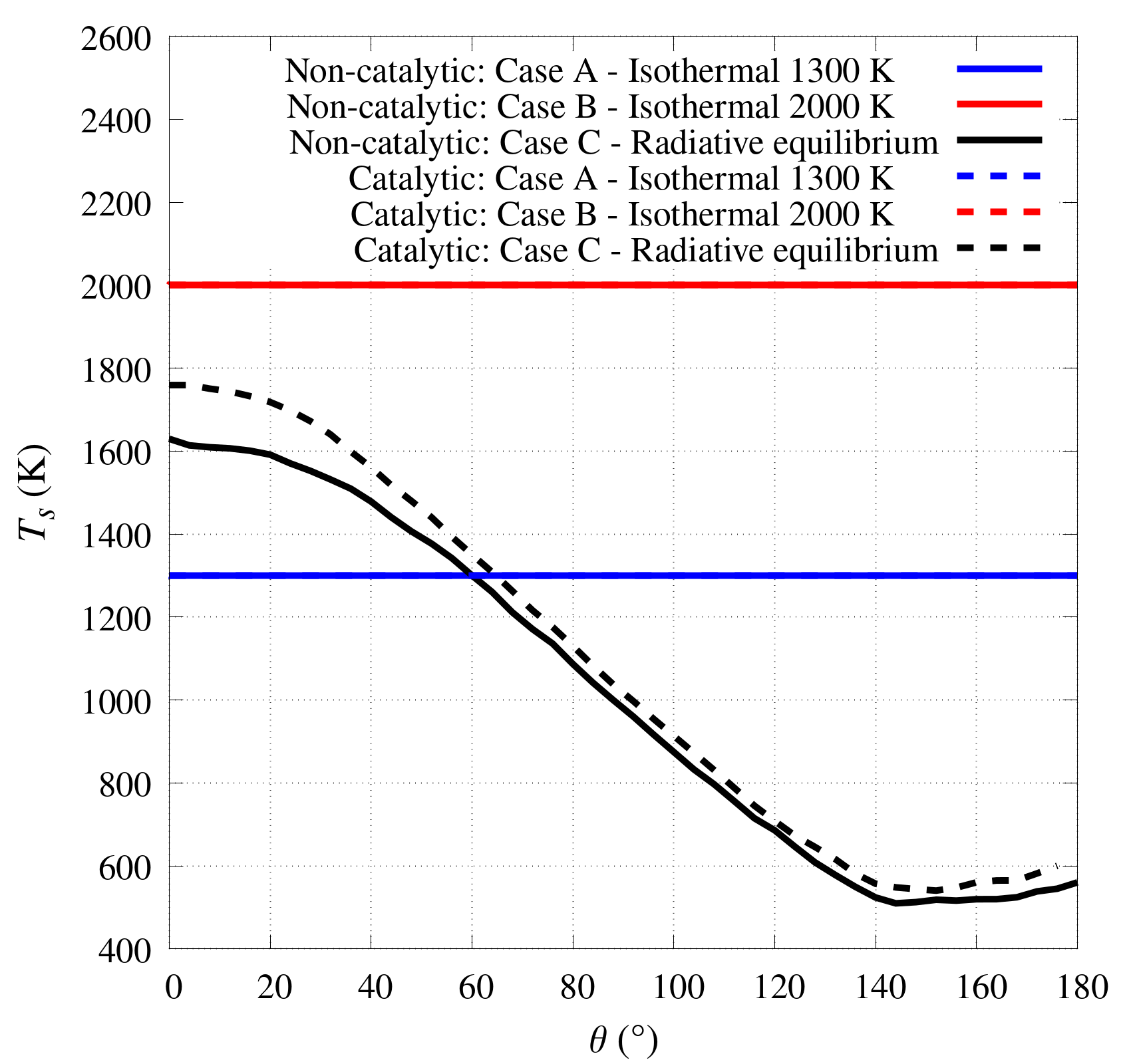}
            \label{fig:surftemp2}
        }
        \caption{Effect of boundary conditions on surface properties at different surface catalycity.}
        \label{fig:bc_standard}
    \end{figure}

    Figure \ref{fig:heatflux2} presents the computed $q_{in}$ for Cases A, B, and C as a function of the surface angle $\theta$ measured from the stagnation point ($\theta = 0^\circ$). Solid lines represent non-catalytic surfaces, while dashed lines indicate catalytic surfaces. The blue, red, and black lines represent Cases A, B, and C, respectively. For all cases, the $q_{in}$ is maximized at the stagnation point. This phenomenon is attributed to the maximal deceleration of the freestream flow at the stagnation point, resulting in the maximum conversion of kinetic energy into thermal energy. This causes elevated gas temperature and density near the stagnation point. Consequently, a greater number of high-energy particles collide with the surface for heat transfer into the surface. As $\theta$ increases, $q_{in}$ decreases due to the reduction in shock strength and the subsequent decline in post-shock temperature and density. Catalytic surfaces consistently exhibit higher $q_{in}$ compared to their non-catalytic counterparts. This phenomenon is attributed to the additional heat load introduced by $q_{chem,in}$. The $T_s$ distribution for each case is shown in Figure \ref{fig:surftemp2}. As expected, Cases A and B maintain a uniform $T_s$ due to the isothermal BC. In contrast, Case C yields a spatially varying $T_s$, with the highest temperature at the stagnation point and a gradual decrease with increasing $\theta$, which follows the local $q_{in}$ variations across the cylinder. A slight increase in $T_s$ beyond $\theta \approx 140^\circ$ is observed due to wake recirculation, where energetic particles re-impinge on the surface, consistent with previous findings in the literature.\cite{Wright2006, Lofthouse2008} Additionally, using the radiative equilibrium BC (Case C), the catalytic surface shows higher $T_s$ than the non-catalytic case due to the increased heat input from surface reactions.
    
\begin{figure}[h!]
        \centering
        \includegraphics[width=0.475\textwidth]{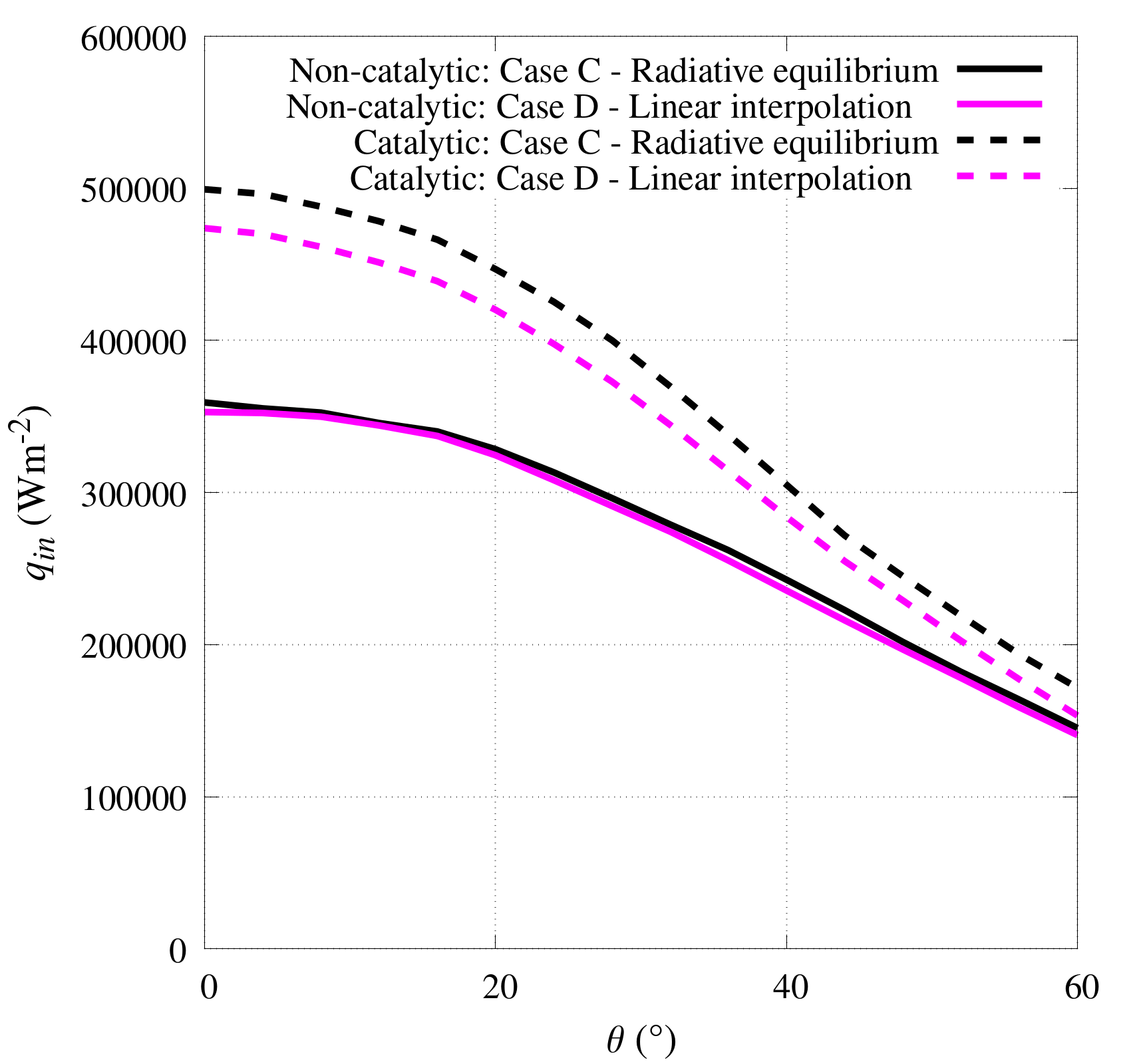}
        \caption{\label{fig:heatflux3}$q_{in}$ as a function of $\theta$ at different surface catalycity.}
    \end{figure}

    If $T_s$ and $q_{in}$ are in perfect linear correlation, the linear interpolation of two isothermal results (Case D) should give the same result as radiative equilibrium BC (Case C). To evaluate this, the results of Case D are compared with Case C in Figure \ref{fig:heatflux3}. The linear interpolation is performed over the $0^\circ \leq \theta \leq 60^\circ$ range, where the predicted $T_s$ for Case C lies within the temperature range of 1300 K to 2000 K. Results for Case D are represented by magenta lines. For the non-catalytic case, the interpolated $q_{in}$ from Case D closely matches that of Case C, with a maximum deviation of only 3.2\% across the considered $\theta$ range. This finding indicates that linear interpolation may provide a reasonable approximation because the correlation between $T_s$ and $q_{in}$ is nearly linear in the absence of surface reactions. However, discrepancies are observed for the catalytic surface. Case D for the catalytic surface underpredicts $q_{in}$ relative to Case C by as much as 10.7\%. This deviation indicates that the correlation between $T_s$ and $q_{in}$ becomes non-linear when surface reactions are involved.

\clearpage
    
\clearpage

    \subsection{Correlation between heat flux and surface temperature}
    \subsubsection{Pearson correlation coefficient analysis}

    To further investigate the relationship between the local $T_s$ and the $q_{in}$, a correlation analysis is conducted. This analysis involves performing DSMC simulations with radiative equilibrium BCs of $\epsilon$ varying from 0.3 to 0.9 in increments of 0.1. The resulting values of $q_{in}$ and $T_s$ for each $\epsilon$ are utilized to calculate the Pearson correlation coefficient, $r(q_{in},T_s)$, which can be defined as,
    
    \begin{equation}\label{eqn:correlation} 
       r(q_{in},T_s) = \frac{\textnormal{cov}(q_{in},T_s)}{\sigma_{q_{in}}\sigma_{T_s}},
    \end{equation}
    
    \noindent where $\textnormal{cov}(q_{in},T_s)$ denotes the covariance, and $\sigma_{q_{in}}$ and $\sigma_{T_s}$ are the respective standard deviations. By definition, $r(q_{in},T_s)=-1$ corresponds to a perfect negative linear correlation, whereas $r(q_{in},T_s)=+1$ indicates a perfect positive linear correlation. Any intermediate value of $-1 < r(q_{in},T_s) < +1$ can be considered a deviation from a perfect linear correlation. In this study, this is referred to as 'non-linear correlation' for simplicity. This analysis is conducted under both non-catalytic and catalytic conditions. 

\begin{figure}[ht!]
        \centering
        \subfigure[$r(q_{in},T_s)$ as a function of $\theta$.]{
            \includegraphics[width=0.6\textwidth]{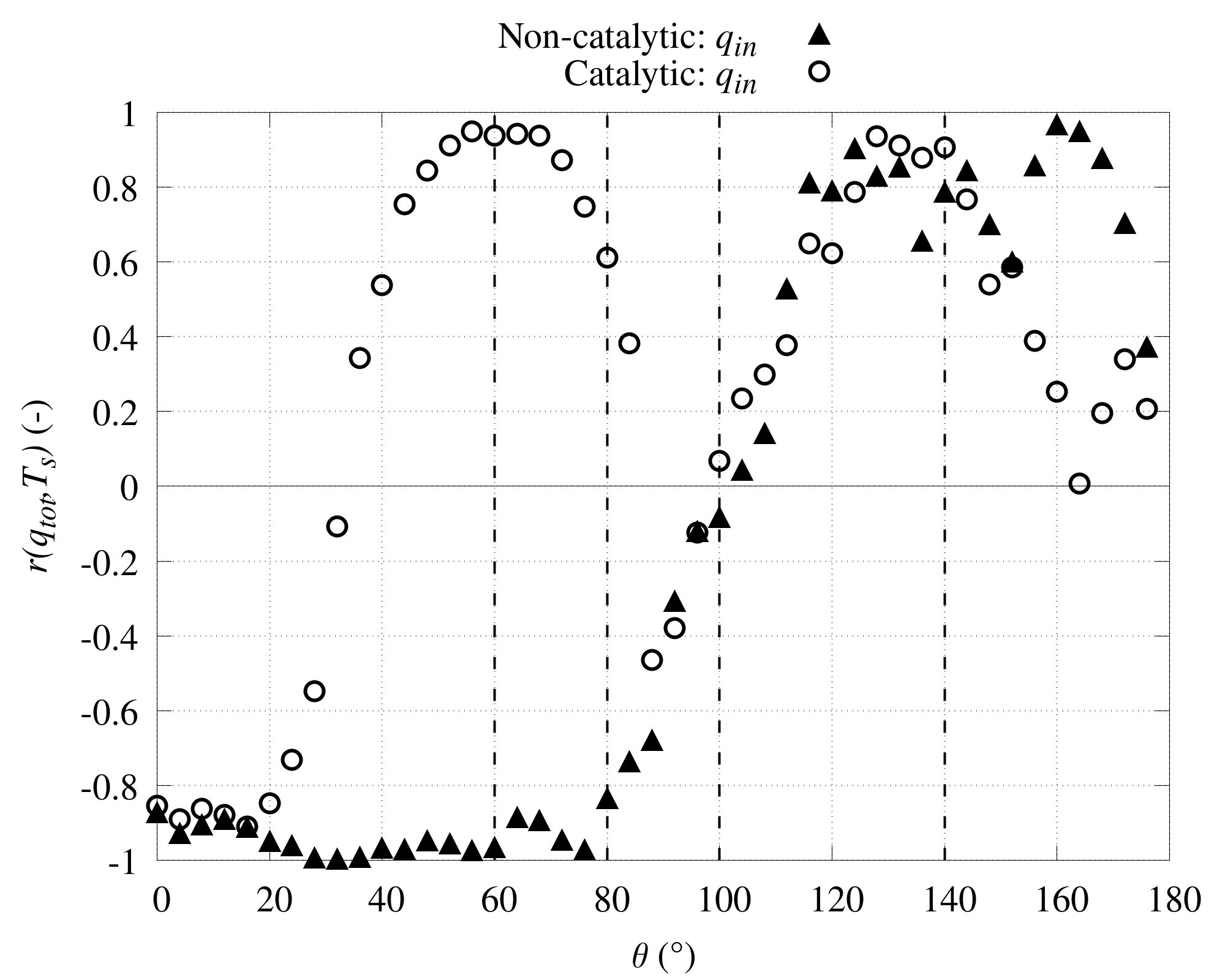}
            \label{fig:correlation1}
        }
        \subfigure[$r(q_{conv,in},T_s)$ and $r(q_{chem,in},T_s)$ as a function of $\theta$.]{
            \includegraphics[width=0.6\textwidth]{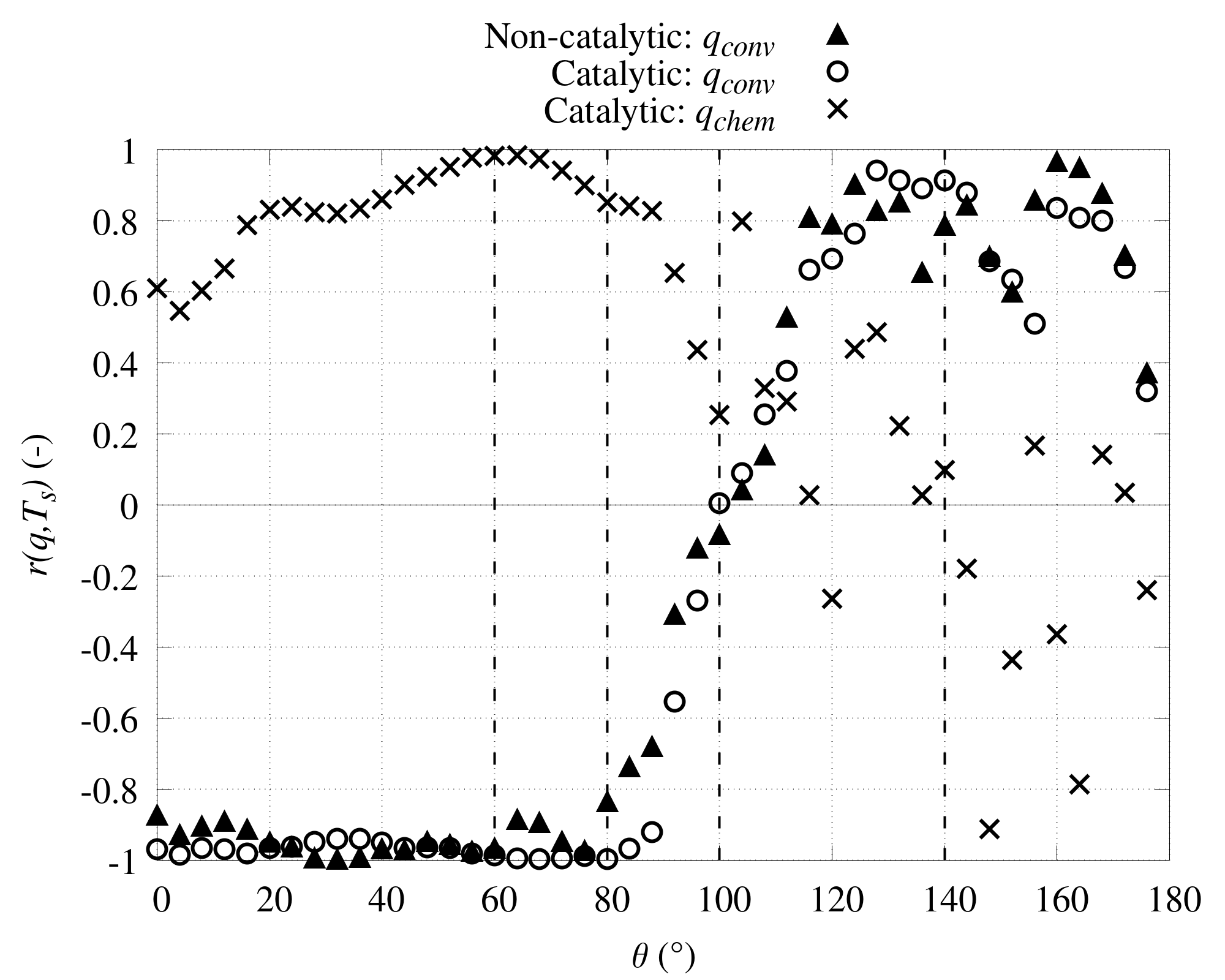}
            \label{fig:correlation2}
        }
        \caption{Correlation coefficient at different surface catalycity.}\label{fig:correlation}
    \end{figure}

    Figure \ref{fig:correlation1} presents the correlation coefficient $r(q_{in},T_s)$ as a function of $\theta$. Solid triangular symbols represent non-catalytic cases, while open circular symbols denote catalytic cases. Focusing on the range $0^\circ \leq \theta \leq 60^\circ$ shown in Figure \ref{fig:heatflux3}, the distinct behavior of $r(q_{in},T_s)$ between non-catalytic and catalytic surfaces is evident. For the non-catalytic case, $r(q_{in},T_s)\approx -1$ indicates a strong negative linear correlation between $T_s$ and $q_{in}$. This trend is consistent with the findings of Yang et al., who report a similar correlation of the two parameters on the forebody of a non-catalytic blunt-headed cone in re-entry conditions \cite{Yang2022}. In contrast, $r(q_{in},T_s)$ of the catalytic case diverges from that of the non-catalytic case. Between $0^\circ \leq \theta \leq 60^\circ$, $r(q_{in},T_s)$ shows a sharp increase from $-1$ to $+1$. This indicates that the two parameters lose their linear correlation at $0^\circ \leq \theta \leq 60^\circ$. 

    To uncover the origin of the differences, $q_{in}$ is decomposed into its components, $q_{conv,in}$ and $q_{chem,in}$, which are evaluated separately for their correlation with $T_s$. The results are shown in Figure \ref{fig:correlation2}. The $r(q_{conv,in},T_s)$ for the non-catalytic surface is represented by solid triangular symbols. These symbols correspond exactly to those of Figure \ref{fig:correlation1}, as $q_{conv,in}$ is the only component of $q_{in}$ for a non-catalytic surface. For the catalytic surface, the $r(q_{conv,in},T_s)$ and $r(q_{chem,in},T_s)$ are denoted by empty circular and cross symbols, respectively. The findings show that the $r(q_{conv,in},T_s)$ for catalytic surfaces closely resembles that of the non-catalytic case. Firstly, this suggests that surface catalycity does not significantly alter the $q_{conv,in}$ correlation with $T_s$. Secondly, it demonstrates that $q_{chem,in}$ is responsible for the non-linear correlation between $q_{in}$ and $T_s$ at catalytic surfaces. Therefore, surface reactions can explain the divergence between Case C (radiative equilibrium BC) and Case D (linear interpolation of isothermal BC) for the catalytic surface between $0^\circ \leq \theta \leq 60^\circ$ found in Figure \ref{fig:heatflux3}. 

\clearpage
    
\clearpage

    \subsubsection{Non-linear correlation of chemical heat flux}
\begin{figure}[h!]
        \centering
        \includegraphics[width=0.8\textwidth]{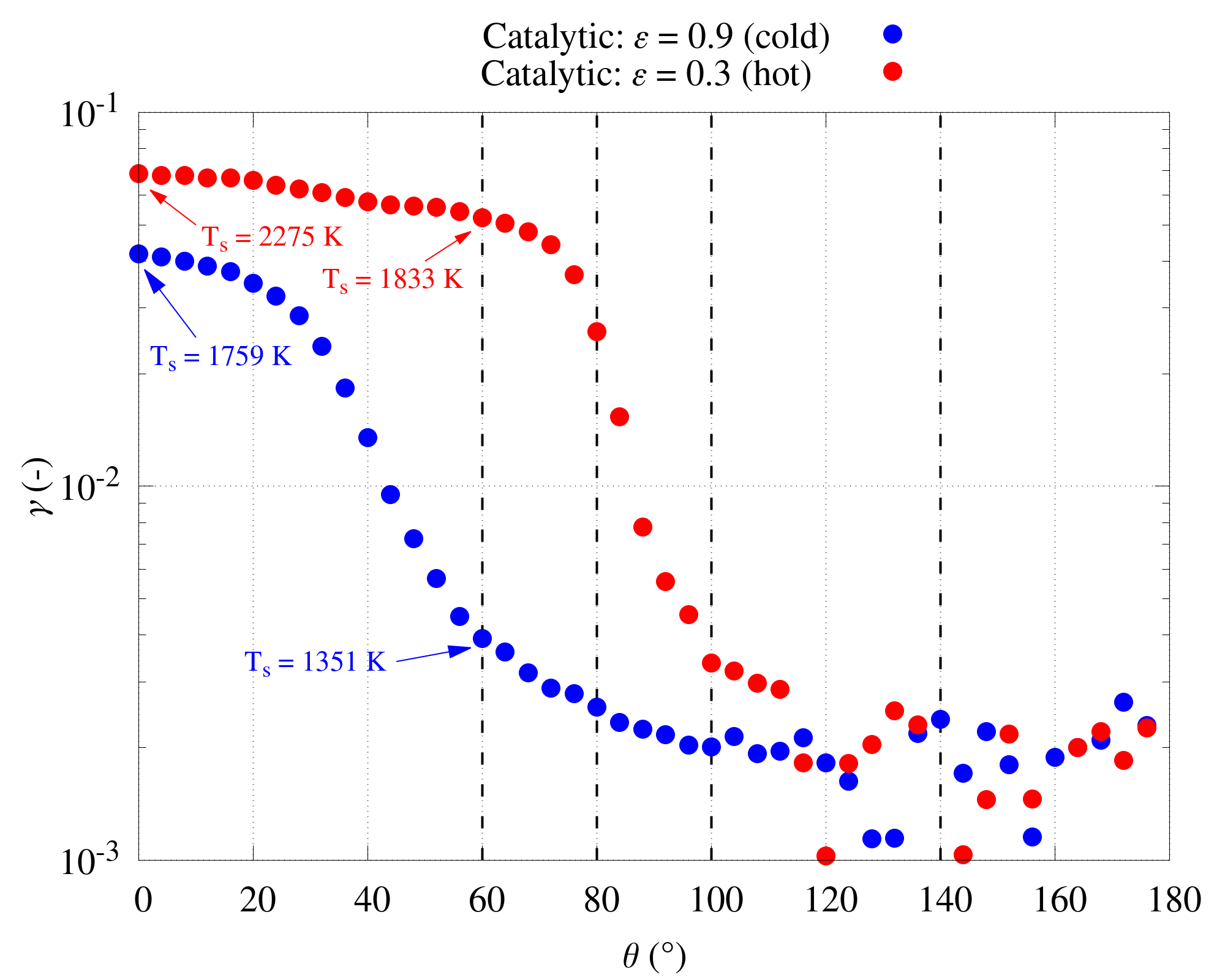}
        \caption{\label{fig:gamma2} $\gamma$ as a function of $\theta$ on catalytic surfaces with different $\epsilon$.}
    \end{figure}

    The correlation between $q_{chem,in}$ and $T_s$ is further analyzed in Figure \ref{fig:gamma2}, which illustrates the distribution of $\gamma$ as a function of $\theta$ for $\epsilon = 0.3$ and $\epsilon = 0.9$. Within the range of interest ($0^\circ \leq \theta \leq 60^\circ$), a lower $\epsilon$ results in a higher $T_s$ for a given $\theta$. For clarity, surfaces with $\epsilon = 0.3$ and $\epsilon = 0.9$ are hereafter referred to as 'hot' and 'cold' surfaces, respectively. The $q_{chem,in}$ is fundamentally dependent on surface catalycity, which is characterized by $\gamma$ with Equation \ref{eqn:gamma}. The 'hot' surface consistently exhibits significantly higher values of $\gamma$ compared to the 'cold' surface. This difference arises due to the strong $T_s$ dependence of heterogeneous recombination reactions, which are integrated into the all-T model utilized in this study. At $T_s$ computed between $0 \leq \theta \leq 60^\circ$ for both cases, the heterogeneous recombination of $O$ is predominantly governed by the ER mechanism, which is exponentially dependent on $T_s$, as previously illustrated in Figure \ref{fig:gamma_all317}. The elevated $\gamma$ with increasing $T_s$ leads to a corresponding increase in $q_{chem,in}$ due to enhanced heterogeneous recombination. This explains why the $r(q_{chem,in}, T_s)$ maintains a positive value between $0^\circ \leq \theta \leq 60^\circ$ in Figure \ref{fig:correlation2}. 
    
\begin{figure}[h!]
        \centering
        \subfigure[$\theta = 0^\circ$.]{
            \includegraphics[width=0.45\textwidth]{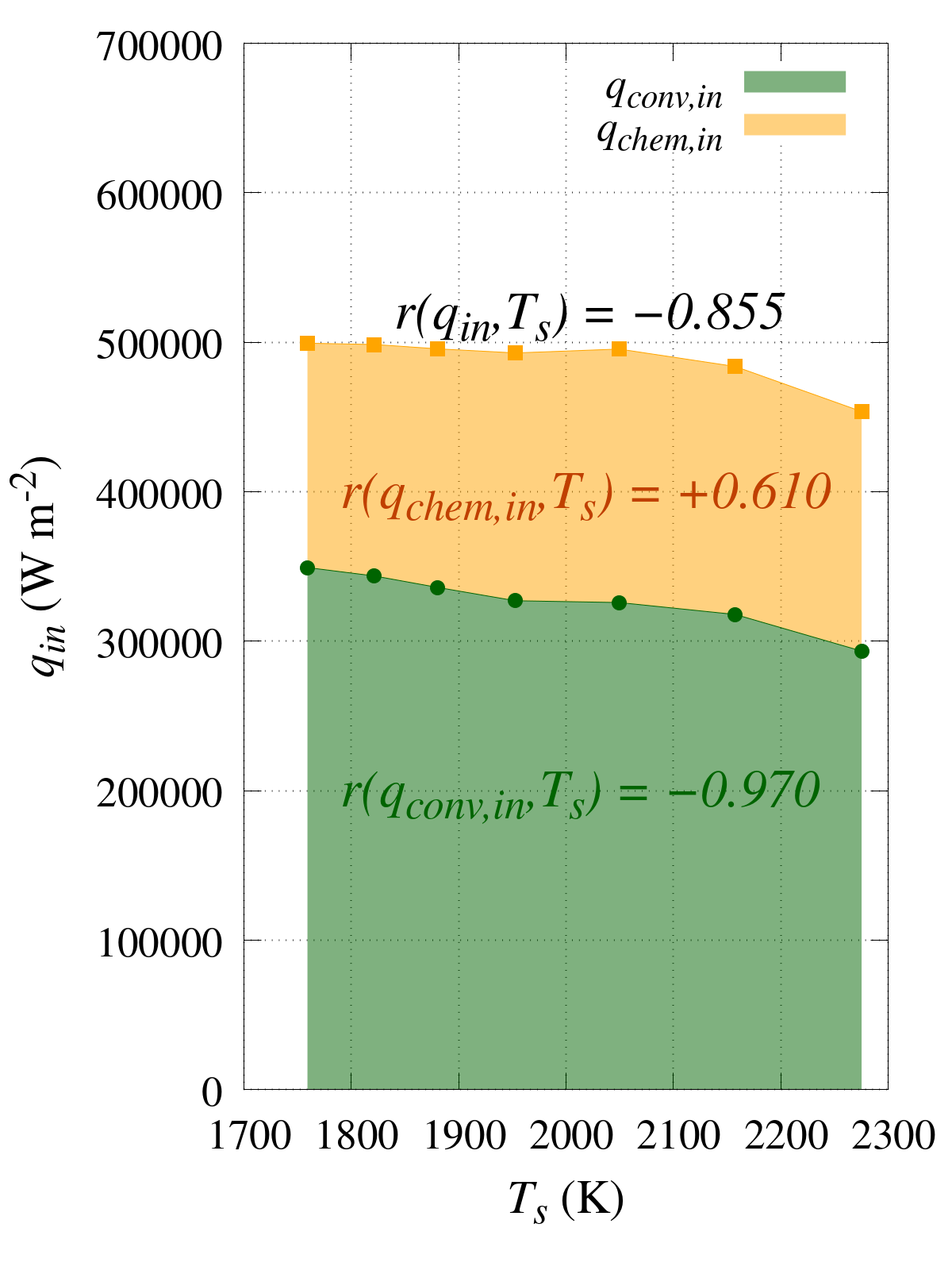}
            \label{fig:heatflux_0}
        }
        \subfigure[$\theta = 32^\circ$.]{
            \includegraphics[width=0.45\textwidth]{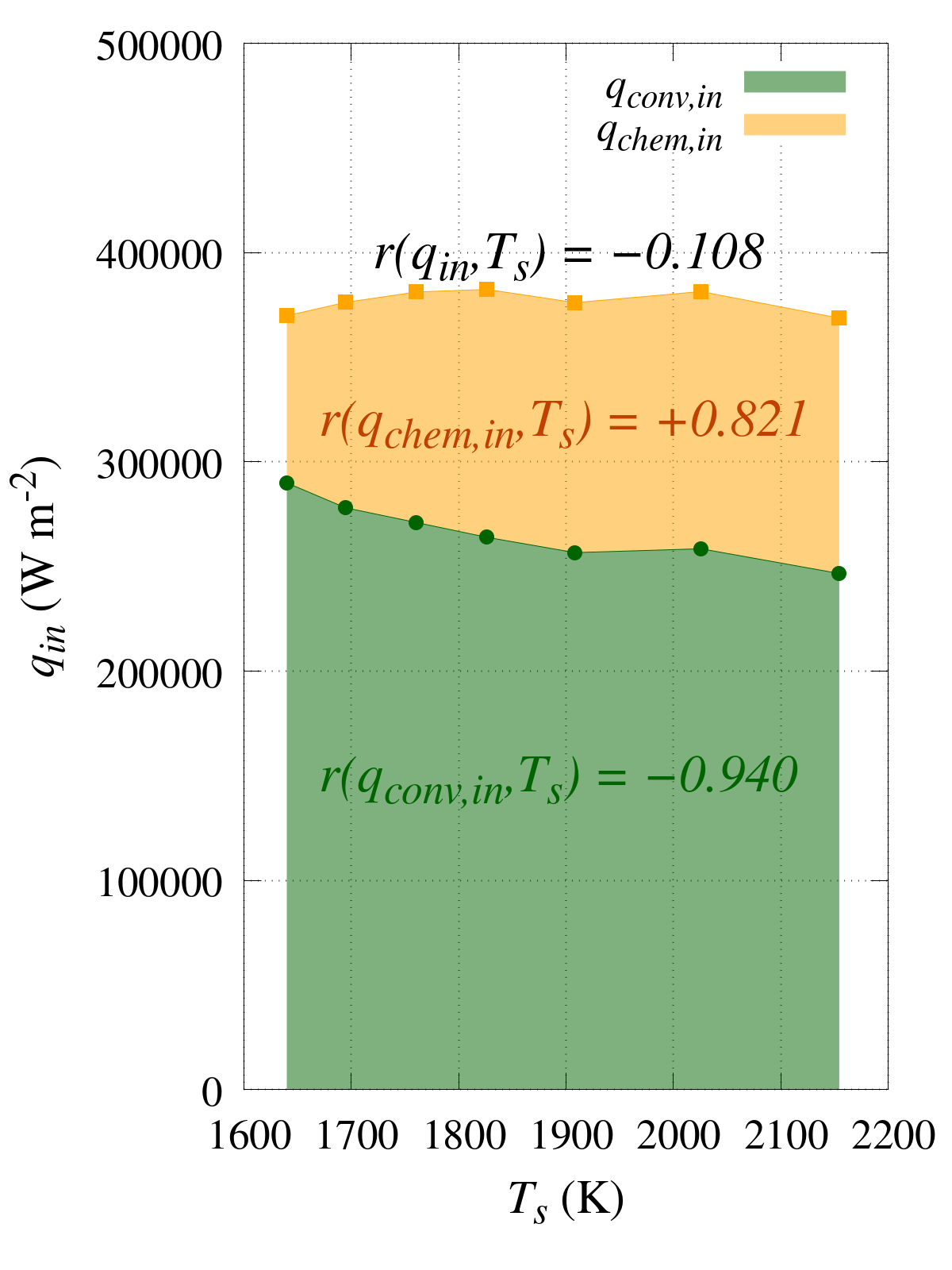}
            \label{fig:heatflux_32}
        }
        \subfigure[$\theta = 60^\circ$.]{
            \includegraphics[width=0.45\textwidth]{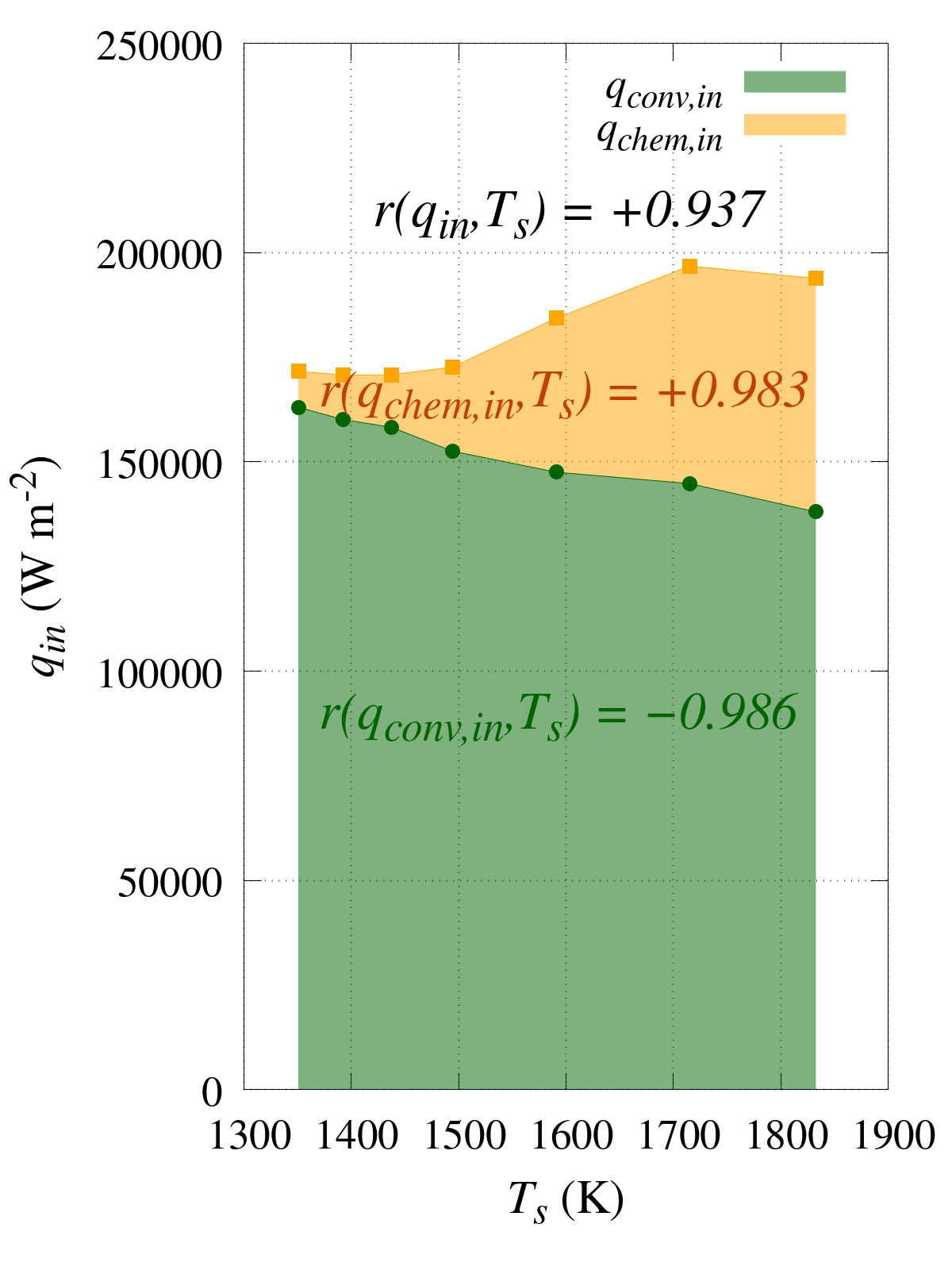}
            \label{fig:heatflux_60}
        }
        \caption{$q_{conv,in}$ and $q_{chem,in}$ as a function of $T_s$ on catalytic surface with different $\epsilon$.}
        \label{fig:heatflux_emiss}
    \end{figure}

    Although $r(q_{chem,in}, T_s)$ remains positive over the interval $0 \leq \theta \leq 60^\circ$, Figure \ref{fig:correlation1} shows that $r(q_{in},T_s)$ transitions from $-1$ at $\theta = 0^\circ$ to $+1$ at $\theta = 60^\circ$. This behavior can be explained by examining the relative magnitudes of $q_{conv,in}$ and $q_{chem,in}$ that contribute to the total $q_{in}$. Figure \ref{fig:heatflux_emiss} illustrates the two heat flux components as functions of $T_s$ measured at $\theta = 0^\circ, 32^\circ,$ and $60^\circ$. The shaded regions in green and yellow represent the relative contributions of $q_{conv,in}$ and $q_{chem,in}$, respectively. The total height of the curve represents $q_{in}$ for each case. The values of $r(q_{in}, T_s)$, $r(q_{conv,in}, T_s)$, and $r(q_{chem,in}, T_s)$ for the three $\theta$ values are consistent with those presented in Figure \ref{fig:correlation}. At each $\theta$ considered, the $r(q_{conv,in}, T_s) \approx -1$. The $q_{conv,in}$ decreases by approximately 15\% from the lowest to the highest $T_s$ at each $\theta$. On the other hand, while $r(q_{chem,in}, T_s)$ remains positive, the extent of the increase in $q_{chem,in}$ with $T_s$ varies significantly at different $\theta$. At $\theta = 0^\circ$, the increase in $q_{chem,in}$ with rising $T_s$ is modest. At $\theta = 32^\circ$, the increase in $q_{chem,in}$ is comparable to the decrease in $q_{conv,in}$. This leads to a $q_{in}$ remaining nearly constant at different $T_s$. At $\theta = 60^\circ$, the increase in $q_{chem,in}$ becomes particularly pronounced, such that $q_{in}$ actually increases with $T_s$. This occurs because the difference in $\gamma$ between 'hot' and 'cold' cases varies along $0^\circ \leq \theta \leq 60^\circ$, as shown in Figure \ref{fig:gamma2}. 
    
\clearpage

\clearpage

    \subsubsection{Non-linear correlation of convective heat flux}
\begin{figure}[h!]
        \centering
        \subfigure[$\bar{j}_{in}$ as a function of $T_s$.]{
            \includegraphics[width=0.475\textwidth]{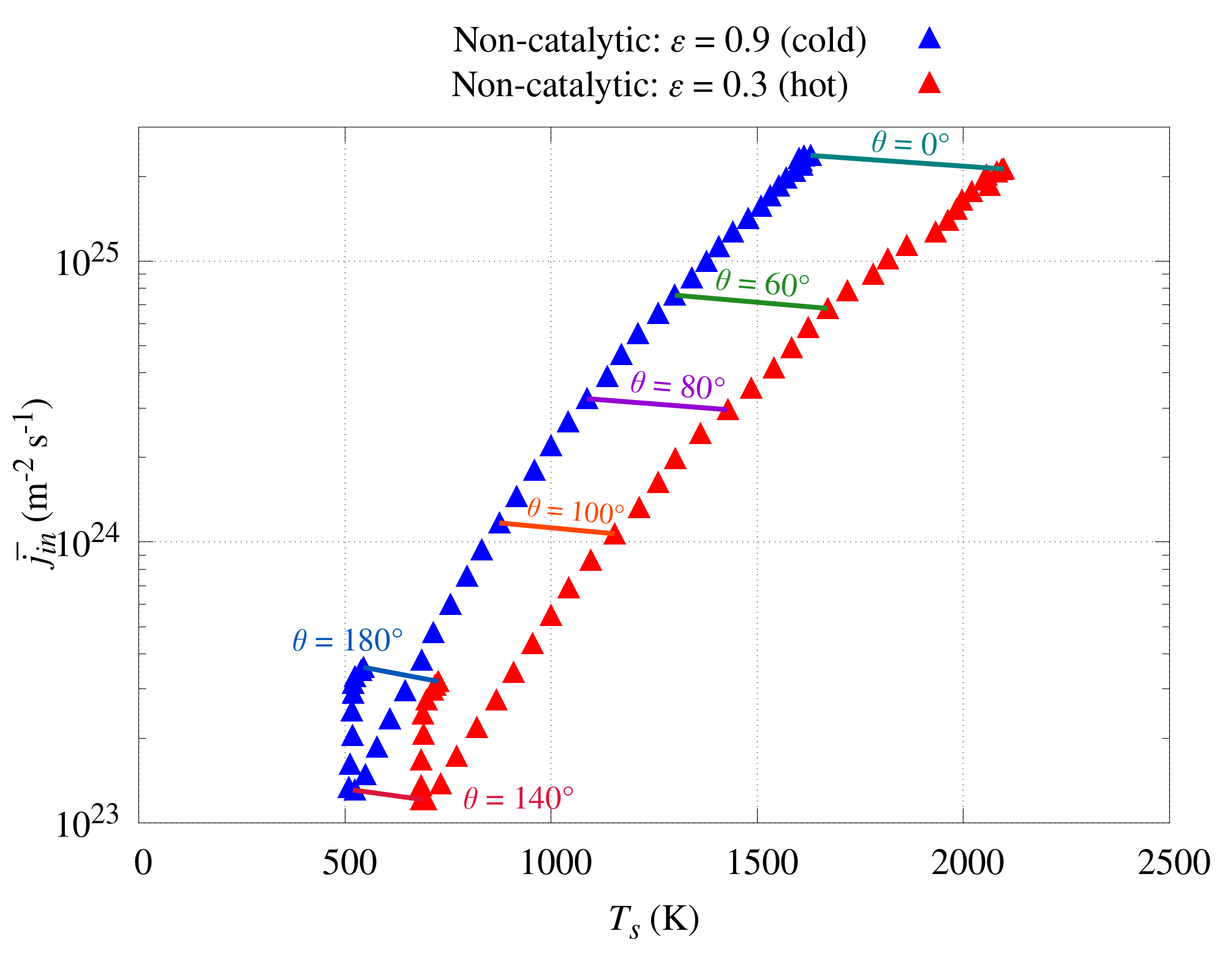}
            \label{fig:perflux_flux}
        }
        \subfigure[$\bar{E}_{in}$ and $\bar{E}_{out}$ as a function of $T_s$.]{
            \includegraphics[width=0.475\textwidth]{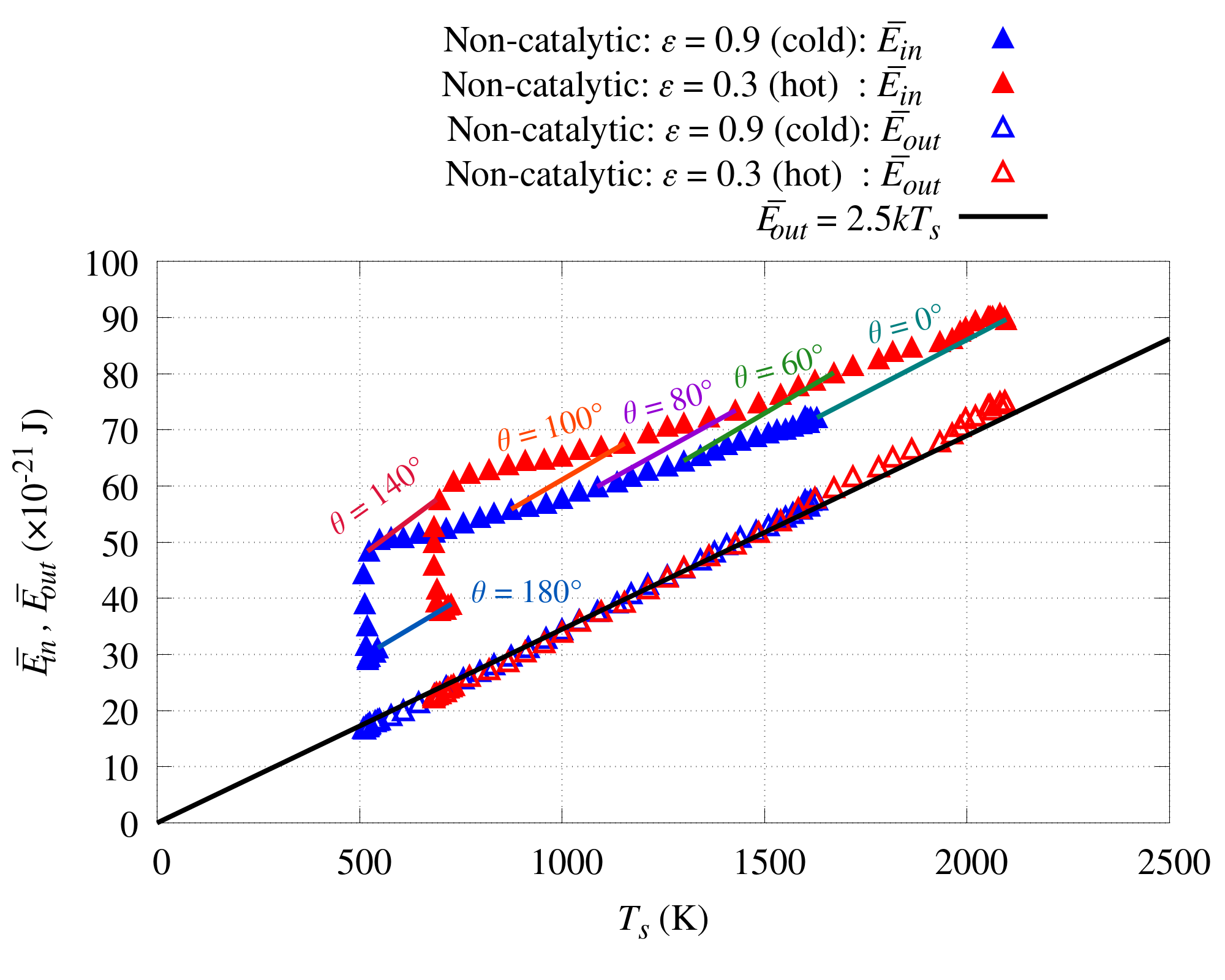}
            \label{fig:perflux_inout}
        }
        \subfigure[$\Delta \bar{E}$ as a function of $T_s$.]{
            \includegraphics[width=0.475\textwidth]{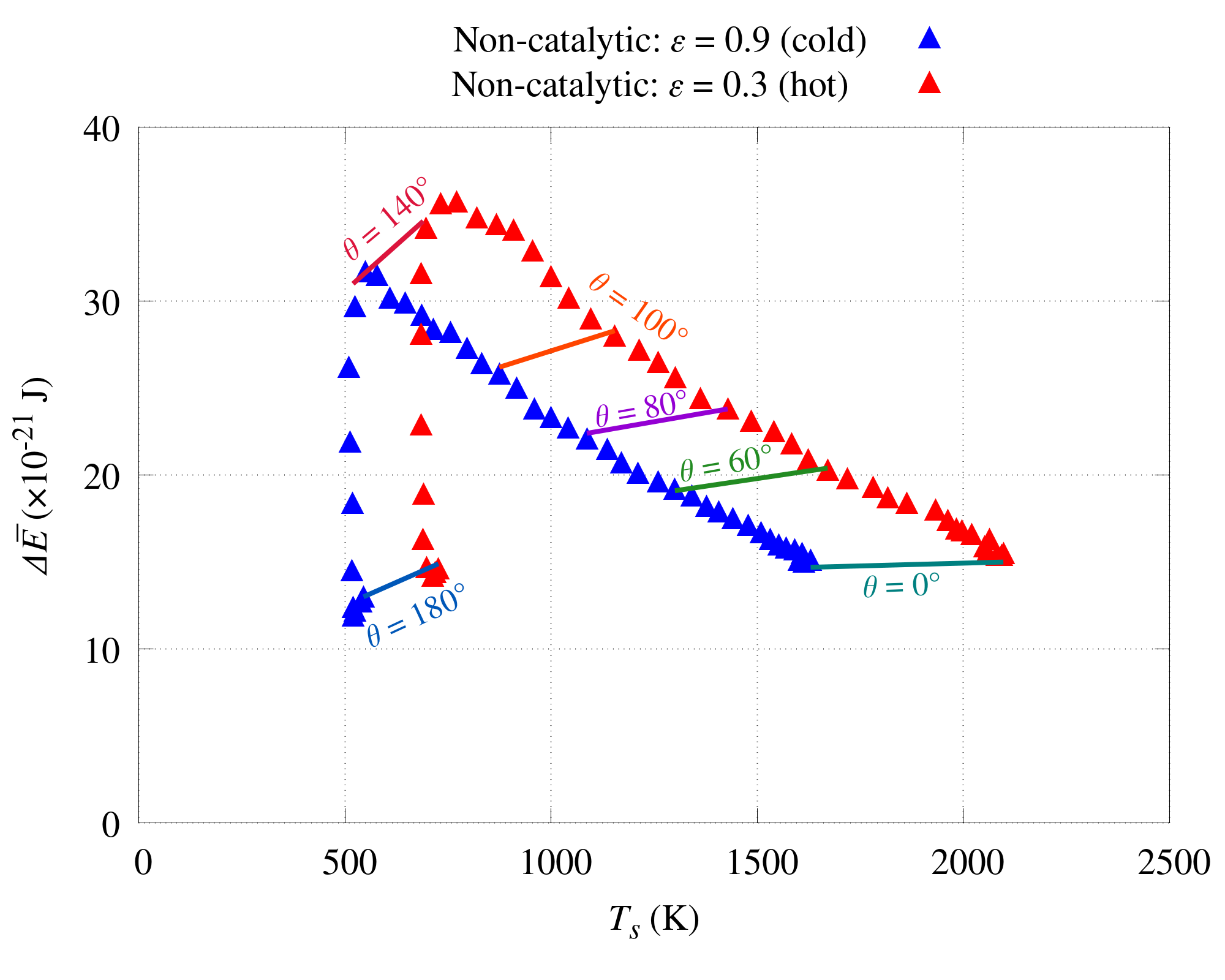}
            \label{fig:perflux_deltaE}
        }
        \caption{Average surface collision parameters on non-catalytic surface with different $\epsilon$.}
        \label{fig:perflux}
    \end{figure}

    Another interesting finding from Figure \ref{fig:correlation2} is the behavior of $r(q_{conv,in},T_s)$ along $\theta$. The $r(q_{conv,in},T_s)$ remains close to $-1$ between $0^\circ \leq \theta \leq 80^\circ$, followed by an increase from $-1$ to $+1$ near the aftbody, specifically between $80^\circ \leq \theta \leq 140^\circ$. The underlying physics governing this trend within the $\theta$ range is analyzed. From a physical standpoint, $q_{conv,in}$ depends on two main factors: the average flux of incident particles, $\bar{j}_{in}$, and the average net energy exchange per gas-surface collision, $\Delta \bar{E} = \bar{E}_{in} - \bar{E}_{out}$. Their product yields $q_{conv,in}$, such that  $q_{conv,in}=\bar{j}_{in} \times \Delta \bar{E}$. Figure \ref{fig:perflux} plots these parameters as functions of $T_s$ for two $\epsilon$, $\epsilon=0.3$ (red) and $\epsilon=0.9$ (blue). The connecting lines mark the key angular positions $\theta = 0^\circ$, $60^\circ$, $80^\circ$, $100^\circ$, $140^\circ$, and $180^\circ$.
    
    Figure \ref{fig:perflux_flux} illustrates $\bar{j}_{in}$ as a function of $T_s$. The $\bar{j}_{in}$ is the largest at $\theta = 0^\circ$ due to the significant compression of gas particles, which maximizes the frequency of gas-surface collisions. Between $0^\circ \leq \theta \leq 140^\circ$, $\bar{j}_{in}$ decreases exponentially while the boundary layer remains attached to the cylinder surface \cite{Park2016}. At $\theta = 140^\circ$, $\bar{j}_{in}$ attains a minimum because the boundary layer separates, resulting in fewer particles impinging on the surface. Beyond this point, flow recirculation increases the likelihood of particle re-impingement and raises $\bar{j}_{in}$. Although it is not immediately apparent due to the logarithmic scale on the y-axis, the $\bar{j}_{in}$ for the 'cold' surface consistently remains 10\% higher than that for the 'hot' surface at a given $\theta$. This phenomenon occurs because gas scattered from a 'cold' surface carries a lower $\bar{E}_{out}$, resulting in higher compressibility and an increased likelihood of these particles re-impinging on the surface, thereby elevating $\bar{j}_{in}$ \cite{Xu2021}. 

    Figure \ref{fig:perflux_inout} illustrates the variation of $\bar{E}_{in}$ (solid symbols) and $\bar{E}_{out}$ (empty symbols) in relation to $T_s$. A linear dependence of $\bar{E}_{out}$ on $T_s$ is observed, which can be expressed as follows:
    
    \begin{equation}\label{eqn:eout}
        \bar{E}_{out} = \frac{5}{2} k_b T_s.
    \end{equation}
    
    \noindent This relationship arises from the choice of $\alpha =1$, where the scattered particles are assumed to undergo complete thermal accommodation with $T_s$. In other words, $\bar{E}_{out}$ depends solely on the local $T_s$. Since $T_s$ decreases with increasing $\theta$ across the 2D cylinder, $\bar{E}_{out}$ would decrease with increasing $\theta$. Between $0^\circ \leq \theta \leq 140^\circ$, the $\bar{E}_{in}$ also decreases with decreasing $T_s$ (or increasing $\theta$). However, the $\bar{E}_{in}$ decreases at a slower rate as a function of local $T_s$, compared to $\bar{E}_{out}$. This phenomenon occurs because, as $\theta$ increases within the attached boundary layer, a greater number of gas-surface collisions involve gas particles that have experienced prior collisions with the surface upstream of the cylinder. These particles would carry energy that corresponds to $\bar{E}_{out}$ expected from prior gas-surface collisions at lower $\theta$. Therefore, unlike $\bar{E}_{out}$, $\bar{E}_{in}$ is indirectly affected by high $T_s$ upstream. The 'memory effect' of $\bar{E}_{in}$ can also explain the higher $\bar{E}_{in}$ observed for a 'hot' surface compared to that of a 'cold' surface computed at a given $\theta$. Beyond $\theta \geq 140^\circ$, the boundary layer separates from the surface. In this region, most gas-surface collisions involve particles originating from the flow recirculation, with reduced thermal 'memory effect' from upstream. This leads to a significant reduction in $\bar{E}_{in}$. 

    Figure \ref{fig:perflux_deltaE} depicts the $\Delta \bar{E}$ as a function of $T_s$. By definition, the $\Delta \bar{E}$ reaches its maximum where the difference between $\bar{E}_{in}$ and $\bar{E}_{out}$ is maximized. This condition occurs at $\theta\approx140^\circ$, where the boundary layer is fully developed and approaches the point of separation. For any fixed $\theta$, the 'hot' surface generates a larger $\Delta \bar{E}$ compared to the 'cold' surface. As previously discussed, this phenomenon can be attributed to the 'memory effect' of gas particles colliding with the surface, whereby a higher upstream $T_s$ results in increased downstream $\bar{E}_{in}$. 
    
    It is observed that, over all $\theta$, 'cold' surfaces consistently yield greater $\bar{j}_{in}$ but lower $\Delta \bar{E}$ compared to 'hot' surfaces. These two parameters exhibit opposing relationships with $T_s$. Given that $q_{conv,in} = \bar{j}_{in} \times \Delta \bar{E}$, the correlation $r(q_{conv,in}, T_s)$ is primarily determined by the dominant contributor among the two terms. Near $\theta = 0^\circ$, the $\bar{j}_{in}$ reaches its maximum value, and it serves as the determining factor for the computed $q_{conv,in}$. Low $T_s$ surface exhibits higher $\bar{j}_{in}$, indicating that they would yield greater $q_{conv,in}$ despite having lower $\Delta \bar{E}$. In other words, $T_s$ and $q_{conv,in}$ are negatively correlated, and $r(q_{conv,in},T_s) \approx -1$ are observed near $\theta = 0^\circ$. As depicted in Figure \ref{fig:correlation2}, this effect may have continued until $\theta = 80^\circ$. Compared to low $\theta$, the magnitude of $\bar{j}_{in}$ is significantly lower at high $\theta$. This leads to the increased importance of $\Delta \bar{E}$ in determining $q_{conv,in}$. Surfaces with higher $T_s$ exhibit greater $\Delta \bar{E}$ across all $\theta$ values, resulting in an increase in $q_{conv,in}$ when the magnitude of $\bar{j}_{in}$ is considerably low. Under the specified freestream and surface conditions, this effect may have occurred between $140^\circ \leq \theta \leq 180^\circ$. This observation explains the trend of $r(q_{conv,in},T_s)\approx +1$ within this $\theta$ range near the aftbody region, as illustrated in Figure \ref{fig:correlation2}. From this explanation, the range $80^\circ \leq \theta \leq 140^\circ$ appears to represent a transition region wherein neither $\bar{j}_{in}$ nor $\Delta \bar{E}$ is dominant in influencing $q_{conv,in}$. This phenomenon can account for the variations in $r(q_{conv,in},T_s)$ from $-1$ to $+1$ within this $\theta$ range. 

\clearpage
    
\clearpage
\section{Conclusion}\label{conclusion}
    This study incorporates a radiative equilibrium BC into the DSMC framework while accounting for surface reactions. Catalytic effects are represented with the FRSC model, which employs an all‑T reaction set for the heterogeneous recombination of $O$ on a $SiO_2$ surface. The radiative equilibrium BC iteratively determines the $T_s$ that satisfies the energy conservation principle at each surface element. The FRSC model and the radiative equilibrium BC are validated through comparisons with analytical solutions. DSMC simulations are conducted for re-entry flow over a 2D cylinder to obtain $T_s$ and $q_{in}$ across the entire surface. Relative to the conventional isothermal BC, the radiative equilibrium BC can capture the local variation in $T_s$. Linear interpolation of results between two independent isothermal BCs deviates from that of radiative equilibrium BC for catalytic surfaces because the relationship between $q_{in}$ and $T_s$ is inherently non‑linear. Analysis over $0 \leq \theta \leq 60^\circ$ shows that the commonly assumed negative linear correlation between $q_{in}$ and $T_s$ is overly simplistic and potentially misleading. Their correlation depends on the combined effects of (1) surface catalycity, which varies with surface material and computed $T_s$; (2) the local surface angle $\theta$; and (3) freestream conditions. Future work will extend the present framework to include additional material response phenomena, such as ablation and surface recession, to further improve numerical predictions of re-entry aerothermodynamics \cite{Padovan2024}.
    
\clearpage

\section*{\label{ack}Acknowledgment}
    This work was supported by Korea Research Institute for defense Technology planning and advancement(KRIT) grant funded by the Korea government(DAPA(Defense Acquisition Program Administration)) (No. KRIT-CT-22-030, Reusable Unmanned Space Vehicle Research Center, 2025) This work was also supported by the National Supercomputing Center with supercomputing resources including technical support(KSC-2025-CRE-0427).

\clearpage

\end{document}